\RequirePackage{fix-cm}
\documentclass{svjour3}

\usepackage[dvipsnames,table]{
  xcolor
}

\usepackage{amsmath,amssymb,amsfonts}
\usepackage[nolist]{acronym}
\usepackage{algorithmic}
\usepackage{colortbl}
\usepackage{cellspace}
\usepackage{enumitem}
\usepackage{pdflscape}
\usepackage{graphbox}
\usepackage{textcomp}
\usepackage[hidelinks]{hyperref}
\usepackage{multirow}
\usepackage{natbib}
\usepackage{booktabs}
\usepackage{csquotes}
\usepackage[utf8]{inputenc}
\usepackage{pgfplots}
\usepackage[many]{tcolorbox}
\usepackage{fancybox}
\usepackage[super]{nth}
\usepackage{subcaption}
\usepackage{framed}
\usepackage{xspace}
\usepackage{venndiagram}
\usepackage{pifont}

\newacro{is}[IS]{Implementation Science}
\newacro{ip}[IP]{intellectual property}
\newacro{aec}[AEC]{Artifact Evaluation Committee}
\newacro{ase}[ASE]{the \emph{International Conference on Automated Software Engineering}}
\newacro{acm}[ACM]{Association for Computing Machinery}
\newacro{ieee}[IEEE]{Institute of Electrical and Electronics Engineers}
\newacro{icse}[ICSE]{the \emph{International Conference on Software Engineering}}
\newacro{cra}[CRA]{Computing Research Association}
\newacro{doi}[DoI]{Diffusion of Innovations}
\newacro{ec}[EC]{European Commission}
\newacro{epsrc}[EPSRC]{Engineering and Physical Sciences Research Council}
\newacro{emse}[EMSE]{\emph{Empirical Software Engineering}}
\newacro{fse}[FSE]{the \emph{Joint European Software Engineering Conference and Symposium on the Foundations of Software Engineering}}
\newacro{msr}[MSR]{the \emph{International Conference on Mining Software Repositories}}
\newacro{nsf}[NSF]{National Science Foundation}
\newacro{nih}[NIH]{National Institutes of Health}
\acrodefplural{aec}[AECs]{Artifact Evaluation Committees}

\DeclareUnicodeCharacter{2212}{−}
\usepgfplotslibrary{groupplots,dateplot}
\usetikzlibrary{patterns,shapes.arrows}
\pgfplotsset{compat=newest}

\makeatletter
\def\cl@chapter{\@elt {theorem}}
\makeatother

\usepackage[capitalize]{cleveref}
\crefname{section}{Sect.}{sections}
\Crefname{section}{Section}{Sections}

\newcommand{\challengeref}[1]{\textbf{(#1)}}

\usepackage{tabularx}
    \newcolumntype{L}{>{\raggedright\arraybackslash}X}
    \newcolumntype{s}{>{\raggedright\arraybackslash}m{2.8cm}}

\newcommand{\presponse}[2]{\enquote{#2} -- P#1}

\newenvironment{shortlist}
{\renewcommand{\item}{\renewcommand{\item}{\unskip\space\textbullet~}}}
{}

\newcommand{\themerow}[4]{\textbf{\emph{#1}:} #2 & \presponse{#3}{#4}\\[1ex]}
\newcommand{\gthemerow}[5]{#1 & \textbf{\emph{#2}:} #3 & \presponse{#4}{#5}\\[1ex]}

\newcommand{\gthemerowx}[4]{#1 & \textbf{\emph{#2}:} #3 & #4\\[1ex]}
\newcommand{\themeheader}{
  \renewcommand\arraystretch{1.6}
  \setlength{\aboverulesep}{0pt}
  \setlength{\belowrulesep}{0pt}
  \rowcolors{2}{white}{lightgray!30}
  \begin{tabularx}{\linewidth}{p{0.45\textwidth}X}
  \toprule
  \textbf{Title and description} & \textbf{Representative quote} \\[.5ex]
  \midrule}
\newcommand{\titledthemeheader}[1]{
  \rowcolors{2}{gray!25}{white}
  \begin{tabularx}{\linewidth}{p{0.4\textwidth}X}
  \rowcolor{lightgray!50}
  \toprule
  \multicolumn{2}{l}{\textbf{#1}}\\
  \midrule
  \textbf{Title and description} & \textbf{Representative quote(s)} \\
  \midrule}
\newcommand{\themefooter}{\bottomrule\end{tabularx}}

\usepackage{amsmath}
\newcommand{\gcreator}{$\blacksquare$}
\newcommand{\gprocessorg}{$\spadesuit$}
\newcommand{\gadvisor}{$\clubsuit$}
\newcommand{\guser}{$\blacktriangle$}
\newcommand{\greviewer}{$\bigstar$}
\newcommand{\gcommunity}{$\blacklozenge$}

\newcommand{\artifacturl}{\url{https://doi.org/10.5281/zenodo.4737346}\xspace}

\newtcolorbox{insights}[2][]{%
  title={Insights for RQ1},
  title filled,
  minipage boxed title*=-5cm,
  colbacktitle=white,
  coltitle=black,
  fonttitle=\bfseries,
  halign title=flush center,
  enhanced,
  attach boxed title to top center={yshift=-2.5mm},
  frame empty,
  borderline={0.3mm}{0mm}{solid},
  boxrule=0.2mm,
  boxsep=1mm,
  top=4mm,
  bottom=2mm,
  sharp corners=all,
  drop fuzzy shadow,
  fonttitle=\bfseries,
  title=#2,#1}

\pgfplotsset{compat=newest}
\usepgfplotslibrary{groupplots}
\usepgfplotslibrary{dateplot}

{\makeatletter
 \gdef\xxxmark{%
   \expandafter\ifx\csname @mpargs\endcsname\relax 
     \expandafter\ifx\csname @captype\endcsname\relax 
       \marginpar{\textcolor{red}{xxx~}}
     \else
       \textcolor{red}{xxx~}
     \fi
   \else
     \textcolor{red}{xxx~}
   \fi}
 \gdef\xxx{\@ifnextchar[\xxx@lab\xxx@nolab}
 \long\gdef\xxx@lab[#1]#2{{\bf [\xxxmark \textcolor{red}{#2} ---{\sc #1}]}}
 \long\gdef\xxx@nolab#1{{\bf [\xxxmark \textcolor{red}{#1}]}}
}

\makeatletter
\newcommand{\bianca}{\renewcommand\NAT@open{[}\renewcommand\NAT@close{]}}
\makeatother
\newcommand*\sqcitep[1]{{\bianca\citep{#1}}}

\journalname{Empirical Software Engineering}
\begin{document}
\title{Understanding and Improving Artifact Sharing in Software Engineering Research}
\author{Christopher S. Timperley
        \and Lauren Herckis
        \and Claire {Le Goues}
        \and Michael Hilton}

\institute{C. S. Timperley, M. Hilton, and C. {Le Goues}, \at School of Computer Science, Carnegie Mellon University, USA\\
    \email{\{ctimperley,mhilton\}@cmu.edu, clegoues@cs.cmu.edu}
\and L. Herckis \at
 Dietrich College of Humanities \& Social Sciences, Carnegie Mellon University, USA\\
  \email{lrhercki@andrew.cmu.edu}
\and Corresponding author: C. S. Timperley (ctimperley@cmu.edu)
}

\date{Accepted: 14 April 2021}

\maketitle

\begin{abstract}

In recent years, many software engineering researchers have begun to include artifacts
alongside their research papers.
Ideally, artifacts, including tools, benchmarks, and data,
support the dissemination of ideas, provide evidence for research claims,
and serve as a starting point for future research.
However, in practice, artifacts suffer from a variety of issues that prevent
the realization of their full potential.

To help the software engineering community realize the full potential of artifacts,
we seek to understand the challenges involved in the creation, sharing, and use of artifacts.
To that end,
we perform a mixed-methods study including a survey of artifacts in software engineering
publications, and an online survey of 153 software engineering researchers.
By analyzing the perspectives of artifact creators, users, and reviewers,
we identify several high-level challenges that affect the quality
of artifacts including mismatched expectations between these groups,
and a lack of sufficient reward for both creators and reviewers.
Using \ac{doi} as an analytical framework, we examine how these challenges
relate to one another, and build an understanding of the factors that affect
the sharing and success of artifacts.
Finally, we make
recommendations to improve the quality of artifacts based on our results
and existing best practices.

\end{abstract}

\section*{Declarations}

\begin{description}
\item[\textbf{Funding:}] Not applicable.
\item[\textbf{Conflicts of interest:}] Not applicable.
\item[\textbf{Availability of data and material:}]
    A replication package for this study, including
    the author survey instrument and results of the publication survey, is available at
    \artifacturl.
\item[\textbf{Code availability}:] Not applicable.
\end{description}

\section*{Compliance with Ethical Standards}

Our study included an online survey of software engineering researchers.
We received approval for our survey from Carnegie Mellon University's
Institutional Review Board.
We have include our informed consent form as part of the additional materials
for this study. 

\acresetall

\section{Introduction}

Artifacts, in the form of tools, benchmarks, data, and more,
play an integral role in software engineering research.
Tools provide tangible, concrete implementations of abstract
concepts and ideas that can be shared, studied, and tested.
Benchmarks are the means by which we evaluate and compare
the implementations of our abstract concepts and ideas,
and serve as a yardstick for measuring
progress in a field.
Datasets and scripts are used to conduct experiments, test hypotheses,
and uncover new insights.
Ultimately, all of the claims that we make are with respect to these
artifacts. Since ideas and competing thoughts cannot be tested quantitatively,
we have to test our hypotheses on concrete implementations of those abstract ideas,
however imperfect.

Artifacts provide rich benefits to the research community.
They allow independent replication experiments to be performed,
enrich the technical understanding of an associated research paper,
and allow others to repurpose, reuse, and extend previous work.
However, for an artifact to be useable by other researchers, it should be complete, structured, and well documented.  This can involve significant effort on the part of authors.
Unfortunately, recent work suggests that many artifacts suffer from a variety of issues
(e.g., lacking documentation and unstated dependencies)
that prevent them from being reused, extended, and replicated by
others~\citep{Collberg15,Collberg16}.
The controversy~\citep{examining-Reproducibility} around this work suggests
that there are no clear standards within the community as to how to
create and evaluate artifacts.

In recognition of the importance of high-quality artifacts,
several software engineering venues have introduced a formal
artifact review and badging process that authors may optionally use.
These processes allow the claims of artifacts to be assessed,
uncover potential usability issues that may be experienced by others,
and provide a signal about the quality of an artifact to the community in the form
of a badge.
However, the implementation of these processes has been met with both
praise and criticism
from members of the community~\citep{beller-aec-blog,Shriram13}.
Given the importance of artifacts to scientific progress within
software engineering research, it is vital that artifacts are
shared and valued by the community, and that researchers are able to
unlock the full potential of artifacts.

In this paper, we report the results of a mixed-methods study to
better understand how the community perceives, creates, uses, shares,
and reviews artifacts, and the challenges
that impede the sharing of high-quality artifacts.
We perform a statistical analysis of recent publications in software engineering
venues, and conduct an online survey of 153 authors of
accepted papers at those venues, including both qualitative and quantitative components.
Using \ac{doi}~\citep{rogers2010diffusion} as a
framework, we perform a secondary analysis of these methodological components,
identify and explore subtle relationships between findings,
and provide a basis for making recommendations.
Finally, we use principles from \ac{is} to provide actionable
recommendations to specific subpopulations based on the results of our analysis.

We find that artifacts are both valued by the community and shared widely:
Almost two thirds of all research-track papers published at \ac{icse},
\ac{fse}, \ac{ase}, and \ac{emse} between 2014 and 2018 provide an accompanying artifact.
From the results of our author survey and publication survey, we identify a number
of high-level challenges that affect the creation, sharing, use, and review
of artifacts.
These challenges include, among others, a perception that the effort required to create and
share artifacts is not worth it, a lack of community standards and guidelines around
artifacts and how they should be reviewed, and the need for creators to provide ongoing
maintenance.

We show that issues associated with the creation, sharing, use, and review of
artifacts are a product of inadequate communication, social systems (e.g.,
empowerment and reward), effects of time (e.g., \enquote{bitrot}), and
technical aspects of the artifacts themselves (e.g., ease of use).
While the community receives numerous benefits from the sharing of artifacts, we find that
the individuals that create and review those
artifacts perceive that there is little reward for their efforts. Among other challenges, mismatched expectations, misaligned
incentives, and poor communication
between the creators, users, and reviewers of
artifacts lead to suboptimal outcomes and experiences for all involved,
and prevent the full potential of those artifacts from being realized.
We argue
that \acp{aec}, responsible for reviewing
artifacts, are well positioned to tackle many of the issues we have identified, and
to elevate and assure the quality of artifacts.

The main contributions of this paper are as follows:

\begin{itemize}
  \item We conduct a mixed-methods study to understand how artifacts are
  created, shared, used, and reviewed
  (\cref{sec:methodology}).

  \begin{itemize}
  \item In one part of our mixed-methods study, we analyze
    all research-track papers published at four software
    engineering venues between 2014 and 2018 to
    determine the prevalence and availability of artifacts
    (\cref{sec:methodology:publication}).

  \item In parallel, we conduct a survey of 153 authors to understand the perception
    of artifacts and to identify the challenges that researchers face
    when creating, sharing, using, and reviewing artifacts
    (\cref{sec:methodology:survey}).
  \end{itemize}

\item
  We present the integrated results of our individual study components (\cref{sec:results}).

\item We use \ac{doi} as a framework
  to understand how the challenges of creating, sharing,
  using, and reviewing artifacts relate to one another in terms of communication channels,
  social systems, time, and characteristics of the artifacts themselves
  (\cref{sec:secondary}).

\item Based on our findings, we use insights from \ac{is}
  to provide actionable recommendations
  to specific subpopulations (e.g., creators)
  (\cref{sec:recommendations}).
\end{itemize}

We provide the following as a part of our replication package:
our author survey materials and quantitative results,
the results of our publication survey,
and the scripts used to mine publication data and
generate our graphs.

\begin{quote}
\centering
\artifacturl
\end{quote}

\section{Background}
\label{sec:background}

In this section, we introduce the reader to software
engineering research artifacts~(\cref{sec:background:artifacts}),
efforts to formally recognize artifacts~(\cref{sec:background:recognition}),
\acl{doi}~(\cref{sec:background:doi}),
and \acl{is}~(\cref{sec:background:is}).

\subsection{Software Engineering Artifacts}
\label{sec:background:artifacts}

For the purposes of this study, we broadly define an artifact as any external materials
or information provided in conjunction with a research paper via a link.
In practice, this consists of any materials developed by authors and linked to
from a research paper.
For example, this would include replication packages, tools/source code,
companion sites, benchmarks, raw data, curated datasets,
survey instruments and results, mechanised proofs, and more.
The \ac{acm} similarly defines artifacts as
\enquote{a digital object that was either created by the authors to be used
  as part of the study or generated by the experiment itself.
  For example, artifacts can be software systems, scripts used to run
  experiments, input datasets, raw data collected in the experiment, or
  scripts used to analyze results.}~\citep{ACMbadging}.

Authors may choose to share artifacts for a variety of motivations:
For example, artifacts may be shared to allow others to replicate, reproduce
or build upon existing work.
Artifacts may be referred to as replication packages
or laboratory packages~\citep{Shull08}.
Historically, the motivation behind sharing artifacts was to ensure
reproducibility
by providing a means of exactly repeating an experiment to obtain the
same (or a similar) result for the purposes of scrutiny and validation
\citep{Shull02,Shull08,Brooks08,basili1999building}.

\subsection{Recognition of Software Engineering Artifacts}
\label{sec:background:recognition}

Motivated by the lack of attention paid to the
software, models, and specifications (i.e., artifacts)
underlying much of the research within software engineering,
the first \ac{aec} in the software engineering
research community was established at \ac{fse} 2011~\citep{Shriram13}.
The \ac{aec} was tasked with formally evaluating the associated artifacts of
accepted research papers.
Since that first \ac{aec},
\ac{fse} has continued to hold \acp{aec} in most editions of the conference, and,
most recently, \ac{icse}
held its first \ac{aec} in the history of the conference in 2019.
In both their current and original form,
artifacts are optionally evaluated following paper acceptance (i.e., authors must
opt-in), and papers cannot be rejected on the basis of their artifacts.

To promote and reward the formal sharing and review of artifacts, the
\cite{ACMbadging} proposed a set of badges
for research articles containing artifacts in ACM publications:
\emph{Artifacts Available},
\emph{Artifacts Evaluated},
and \emph{Results Validated}.
The badging scheme provides structure to the outcome of the
artifact evaluation process while allowing conferences and journals to
continue to review artifacts as they best see fit.
As of July 2020, both \ac{fse} and \ac{icse} participate in the \ac{acm}'s badging
scheme.
Awarded badges appear on the front page of the paper itself within the
proceedings, and are recorded in the metadata of the \ac{acm}'s Digital Library.

Each badge is considered independently, and a paper may be awarded
all badges if it meets the appropriate criteria.
While most badges are awarded based on review by the AEC, authors technically may request publishers
to award them an \emph{Artifacts Available} badge without the need for formal review
(e.g., ICSE 18, where two papers have badges despite no formal artifact review process).
After a submitted paper has been accepted by the conference, it may be awarded
either an \emph{Artifacts Evaluated: Functional} or an
\emph{Artifacts Evaluated: Reusable} badge
depending on its level of quality
and potential for reuse and repurposing. The \cite{ACMbadging} measures the quality of an
artifact in terms of the extent to which it is \emph{documented},
\emph{consistent},
\emph{complete},
and \emph{exercisable}, according to the following definitions:

\begin{description}
\item[\textbf{Documented:}]
  At minimum, an inventory of artifacts is included, and sufficient description
  provided to enable the artifacts to be exercised.
\item[\textbf{Consistent:}]
  The artifacts are relevant to the associated paper, and contribute in some
  inherent way to the generation of its main results.
\item[\textbf{Complete:}] To the extent possible, all components relevant to the
  paper in question are included. (Proprietary artifacts need not be included.
  If they are required to exercise the package then this should be documented,
  along with instructions on how to obtain them. Proxies for proprietary data
  should be included so as to demonstrate the analysis.)
\item[\textbf{Exercisable:}]
  Included scripts and/or software used to generate the results in the
  associated paper can be successfully executed, and included data can be
  accessed and appropriately manipulated.
\end{description}

Crucially, the \ac{acm} leaves the interpretation of its badging policy and
the implementation of an associated artifact evaluation process to
individual conferences and communities.
Within their artifact review and badging policy~\citep{ACMbadging}, the \ac{acm} states,
\enquote{We believe that it is still too early to establish more specific
guidelines for artifact and replicability review. Indeed, there is sufficient
diversity among the various communities in the computing field that this may
not be desirable at all.}

In addition to badges, several conferences (e.g., \ac{icse}) have introduced
distinguished artifact awards to recognize and reward the creation and
sharing of high-quality artifacts.\footnote{\url{https://2020.icse-conferences.org/info/awards} [Date Accessed: \nth{13} March 2021]}

\subsection{Diffusion of Innovations}
\label{sec:background:doi}

\Ac{doi} is a framework from the social
sciences that seeks to explain how new objects, ideas, and practices spread~\citep{rogers2010diffusion}.
A scientific understanding of how and why new ideas take hold and spread rapidly, or are
briefly acknowledged and then pass into obscurity, is valuable in disciplines
ranging from medicine to information technology to anthropology
(e.g., \citealt{gomez2013study, johns1993constraints, o1998patterns, premkumar1994implementation, wright1995importance}).
In software engineering, \ac{doi} has been used to study
link sharing on stack overflow~\citep{DBLP:conf/msr/GomezCS13},
how to introduce developers to new practices~\citep{green2000successful},
how developers use mobile development platforms~\citep{DBLP:conf/icse/MirandaFSFS14}, and even to
analyze if developers discover new tools on the
toilet~\citep{DBLP:conf/icse/Murphy-HillSSJW19}.

In this paper we use \ac{doi} as an analytical framework to integrate findings drawn from diverse components of our mixed-methods research design, explain relationships between them, and develop a more complete understanding of artifacts in software engineering research.

There are four central elements of \ac{doi} that we apply in this paper.
Below we briefly describe the elements and their meaning in the context of
this paper.

\begin{itemize}
\item An \textbf{innovation} is any novel thing,
idea, procedure, or system.
It need only be perceived as new by the individual
or organization that might adopt it, and may only have one aspect that is
novel. In this paper, we consider individual artifacts as innovations.

\item \textbf{Communication channels} are the various ways that innovations
are distributed from a person of origin to a recipient.
In this paper, we consider artifact links, \acp{aec}, and artifact badges
as communication channels between artifact creators, potential users,
and reviewers.

\item \textbf{Time} plays a significant role in the diffusion of innovations.
  Innovations are not adopted instantly, but instead they spread and must
  remain relevant over time. In this paper, we consider the factors that
  affect the availability and usability of artifacts over time.

\item
All innovation, and all diffusion of innovation, takes place in the context of
a \textbf{social system}. A social system is a set of interrelated units that are engaged
in joint problem solving to accomplish a common goal.
In this paper, we consider the social system to be the software engineering research community
and its evolving set of norms, practices, members, and values, working towards the common
goal of furthering research.

\end{itemize}

\subsection{Implementation Science}
\label{sec:background:is}

\Ac{is} is an empirical approach to understand the factors
that effectively advance research and move research to
practice~\citep{bauer2015introduction}. Functionally, \ac{is} is a series of
iterative processes in which effective practices and policies are identified,
trialed, observed, and improved upon by researchers over the course of many
studies.

\Ac{is} is a relatively new discipline, developed to promote the rapid transformation of
medical research into more effective medical practice and especially to coordinate research and impact practice~\citep{zerhouni2003nih}.
In these diverse fields such as public health~\citep{glasgow1999evaluating},
business~\citep{frambach2002}, and education~\citep{herckis2018passing},
researchers and practitioners have an interest in making results of research
widely available, enabling others to build on these results, and making a rapid
impact on practice. The principles of iterative, system-oriented,
evidence-based improvements embedded in \ac{is} have been demonstrated to facilitate
the diffusion of innovations.

\section{Methodology}
\label{sec:methodology}

\begin{figure}[t]
\centering
\includegraphics[width=1.0\textwidth]{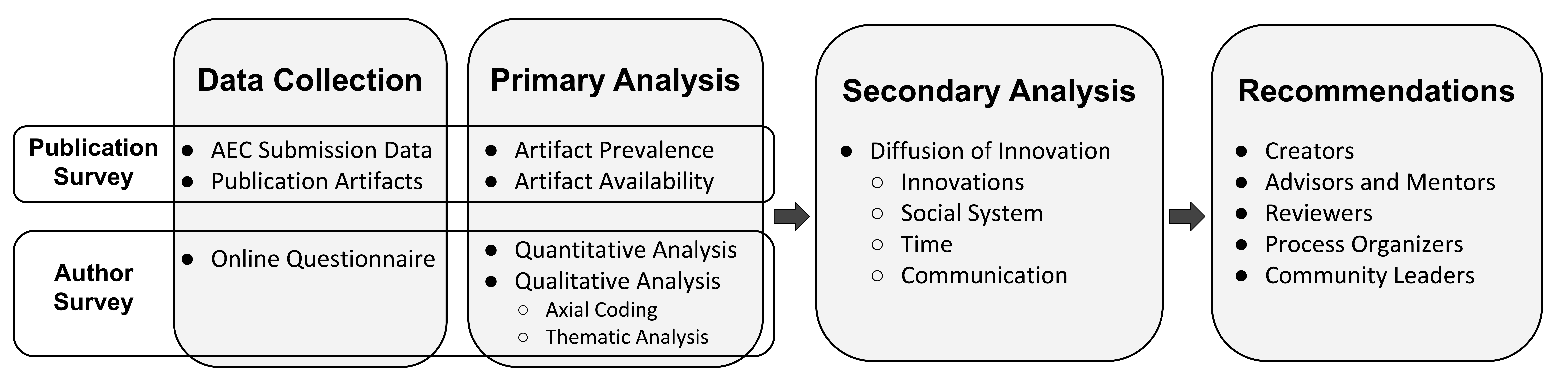}
\caption{An overview of our study methodology.}
\label{fig:methodology}
\end{figure}

In this study, we set out to obtain insights into the values, norms, and practices of
the software engineering research community.
To determine the reasons and extent to which artifacts are created, shared, and used
by researchers, we ask:

\begin{description}
\item[\textbf{RQ1}] \emph{Does the software engineering research community perceive inherent value in artifacts?}
\end{description}

To understand the
challenges that prevent the community from realizing the full potential
of artifacts, we ask:

\begin{description}
\item[\textbf{RQ2}] \emph{What are the challenges of creating, sharing, using,
  and reviewing artifacts?}
\end{description}

We split RQ2 into the following subquestions to understand the challenges associated
with artifacts from the perspective of users, creators, and reviewers.

\begin{description}
\item[\textbf{RQ2.1}] \emph{What challenges are faced by artifact creators?}
\item[\textbf{RQ2.2}] \emph{What challenges are faced by artifact users?}
\item[\textbf{RQ2.3}] \emph{What challenges are faced by artifact reviewers?}
\end{description}

To answer these questions, we used a mixed-methods approach,
outlined in \Cref{fig:methodology},
consisting of
a survey of authors (\cref{sec:methodology:survey}) and a statistical
analysis of publication, submission, and evaluation data related to software
engineering artifacts (\cref{sec:methodology:publication}).
We conduct a secondary analysis using \ac{doi} as an analytical framework
to identify relationships between the challenges
faced by creators, users, and reviewers within the context of the software
engineering research community (\cref{sec:methodology:secondary}).
Finally, we use IS to provide recommendations, targeted at specific subpopulations,
based on our analysis and existing best practices from
the literature (\cref{sec:methodology:recommendations}).

\subsection{Publication Survey}
\label{sec:methodology:publication}

To better understand the prevalence of artifacts within software engineering
research papers, we studied all technical track papers published between 2014
and 2018, inclusive, at three top software engineering conferences
(\ac{icse}, \ac{fse}, \ac{ase})
and one
journal (\ac{emse}).

We first used the DBLP archive\footnote{\url{https://dblp.uni-trier.de} [Date Accessed: March \nth{13} 2021]} to
obtain a list of all technical track papers published between 2014 and 2018
(inclusive) at each of these venues.
We then
downloaded a PDF for each paper and used
PDFx~\citep{pdfx} to transform that PDF
into plaintext, before using regular expressions to find all possible URLs
within each paper.
After finding a list of possible URLs in each paper, we
manually examined each URL to determine if it corresponded to an artifact; in
the case where a paper had no URLs corresponding to artifacts, we manually
inspected the paper to ensure that an artifact URL had not been missed.
In total, we identified 899 artifacts across 1434 papers.
Finally, we determined the availability of each artifact by manually checking that
its associated URL could be accessed at the time of inspection (between
January \nth{29} and February \nth{6}, 2019).

To determine how many papers containing artifacts were submitted to
an \ac{aec} for review, as well as the associated acceptance rate of submissions,
we contacted the \ac{aec} organizers for all conference years within our dataset
that had an \ac{aec} (i.e., \ac{fse} 2015--2018) via email.

\subsection{Author Survey}
\label{sec:methodology:survey}

We designed and distributed an online questionnaire to members of the software engineering research community to identify
the perceived value and challenges of artifact
creation, sharing, use, and review.
We used a survey containing a total of 28 questions, shown in
\Cref{tab:survey-questions}, to probe the
intersecting subpopulations of artifact creators, users, and reviewers. The survey included both selection-based and open-ended questions.
Branching logic was used to identify the subpopulations to which
respondents belonged and show questions that were relevant to those
subpopulations.
For example, respondents who had previously served on an \ac{aec}, or who were
currently serving on an \ac{aec}, were asked questions about that experience;
respondents who had not served on an \ac{aec} were not asked those questions.

\begin{table}[t]
\centering
\renewcommand\arraystretch{1.4}
\setlength{\aboverulesep}{0pt}
\setlength{\belowrulesep}{0pt}
\rowcolors{2}{gray!25}{white}
\begin{tabular}{rp{80mm}p{18mm}}
\toprule
\textbf{\#} & \textbf{Survey Question} & \textbf{RQs} \\
\midrule
  Q1  & Which would you say best describes your primary role? &  \\
  Q2  & Roughly how many publications have you authored? &  \\
  Q3  & Have you ever published a paper that included an artifact? & 2.1 \\
  Q4  & Why have you found that publishing artifacts was not the right choice? & 2.1 \\
  Q5  & Consider your most recently published paper in which you shared an artifact. Select all the reasons below for which you shared that artifact. & 1 \\
  Q6  & Consider up to your three most recent papers published in a software engineering venue. Was there an artifact shared with any of these papers?  & 2.1 \\
  Q7  & Consider up to your three most recently published papers. State all of the reasons below for which you chose not to (or were unable to) create or share an artifact associated with those papers. & 2.1 \\
  Q8  & Consider up to your three most recently published papers. Did you experience any difficulties or challenges when creating artifacts associated with those papers? & 2.1 \\
  Q9 & Have you ever submitted to an artifact evaluation committee? & 2.1 \\
  Q10 & For which of the following reasons have you used an artifact created by someone else? & 1 \\
  Q11 & What challenges have you faced when trying to use an artifact other than your own? & 2.2 \\
  Q12 & What are the most important things that an artifact should have, be, or do? & 1 \\
  Q13 & Have you ever evaluated artifacts as a member of an Artifact Evaluation Committee? & 2.3 \\
  Q14 & List up to three problems or challenges with the artifact evaluation process that you are aware of as a former member of an \ac{aec}. & 2.3 \\
  Q15 & List up to three ways that the artifact evaluation process could be improved. & 2.3 \\
  Q16 & What other positive or negative experiences with artifacts do you think we should know about? & 1, 2.1, 2.2, 2.3 \\
\bottomrule
\end{tabular}
\caption{A list of our survey questions, mapped to their relevant
  research questions.}
\label{tab:survey-questions}
\end{table}

\paragraph{Recruitment}

To obtain an appropriate sample of the authors in our publication dataset,
we identified the subset of authors who had authored at least one technical track
publication at \ac{icse}, \ac{fse}, \ac{ase}, or \ac{emse} in 2018
---
the most recent complete year at the time the survey was conducted. We chose
this approach because we wanted to ensure that survey respondents would be able
to reflect on recent experiences.
Collecting email addresses for each of those authors was a time-consuming manual process:
We first consulted the contact information in the paper, where available,
before using a search engine.
Survey respondents were not paid for taking the survey.

In total, we obtained email addresses for 744 authors, to whom we subsequently
sent the survey.
46 of the 744 survey emails that we distributed failed to deliver.
Of the remaining 698 emails that were delivered, 153 recipients completed the
survey, producing a very strong 22\% response rate that exceeded our expectations.
Demographic information for our participants is provided in \Cref{tab:demographics}.

\begin{table}[t]
\centering
\begin{tabular}{lrr|lrr}
\toprule
\textbf{Role} & \textbf{\#} & \textbf{\%}  & \textbf{Subpopulation} & \textbf{\#} & \textbf{\%}  \\
\cmidrule(l){1-3} \cmidrule{4-6}
Academic Researcher     & 139 & 91 & Creators & 137 & 90 \\  
Industrial Practitioner & 6 & 4 & Users & 129 & 84 \\        
Industrial Researcher   & 8 & 5 & Reviewers & 32 & 21 \\     
\bottomrule
\end{tabular}
\caption{The self-reported role and subpopulations of our 153 survey participants.}
\label{tab:demographics}
\end{table}

\paragraph{Qualitative Analysis}
We analyzed the qualitative components of our survey responses using a
descriptive coding approach \citep{saldana2015coding},
in which responses related to each segment of data is assigned basic labels to
create an inventory of codes.
This process was undertaken
collaboratively by domain experts (Timperley, Hilton, Le~Goues) and a methodologist (Herckis),
after which
adjudication and code mapping were used to refine codes and collapse
categories. Finally, we used an axial coding approach to strategically
organize data and determine which themes were dominant and which less
important, as well as to identify themes that offer opportunities for policy,
process, or practice improvement \citep{charmaz2014constructing}.

Note that the goal of thematic analysis is to identify the full range of \textit{themes}
that characterize some class of experiences. This process entails continued
qualitative data gathering and analysis until analysts have reached thematic
saturation, when no new properties, dimensions, conditions, or consequences can
be identified in the data. A thematic analysis does not tell us how prevalent
each of the experiences represented in the exhaustive inventory of themes might
be, only that they are present in the social context~\citep{saldana2015coding}.

\subsection{Secondary Analysis}
\label{sec:methodology:secondary}

These two parallel studies, the survey of publications and survey of authors,
resulted in a set of primary results that situate artifacts in the broader
context of the software engineering community. We performed a secondary analysis
to contextualize and integrate results of these component studies by applying
the \ac{doi} framework as an analytical tool~\citep{creswell2017designing}.
We consider the results of our primary
analyses in the context of the \ac{doi} framework
allowed us to examine relationships
between various findings.
This analysis results in a robust, descriptive
picture of the landscape of norms and practices, which uniquely positions this
type of research to inform policy recommendations and the creation of
evidence-based guidelines.

\subsection{Recommendations}
\label{sec:methodology:recommendations}

Based on the challenges identified in our primary analysis,
and a broader understanding of the context in which artifacts are created, used, shared, and
reviewed, we make recommendations
that address specific challenges and adhere to IS-based principles.
We assess each recommendation to ensure that it does not exacerbate any
known challenges, is compatible with existing practices, and is likely to be
scalable, sustainable, trialable, and observable \citep{rogers2010diffusion}.
Some of our recommendations are supported by
existing literature, while others are novel and arose directly from the present
research. Each of these recommendations can be implemented, trialed, observed,
and evaluated to determine whether it has achieved its intended outcome.

Recommendations are most effective when they target specific subgroups within a
social system \citep{wolfe1994organizational}. To that end, we tailor our
recommendations to the following subgroups:

\begin{description}
\item[\textbf{Creators:}]
Includes both \emph{primary researchers}, typically students and postdocs,
who are predominantly responsible for creating and maintaining artifacts,
and the \emph{mentors and advisors} of those primary researchers.

\item[\textbf{Primary Researcher:}]
Primary researchers
are extensively involved in most aspects of research projects (e.g.,
writing code, collecting data, running experiments), including the
creation and maintenance of artifacts.
This role often, but not always, falls to students and postdocs.
As those working most closely with artifacts, primary researchers can
take several steps to elevate the quality of their
artifacts, reduce unnecessary work for both themselves and their potential
users, and ensure that they are credited for their work.

\item[\textbf{Research Mentors and Advisors:}] Traditionally, this role belongs
  to the advisor of the primary researcher, but there may be multiple mentors
  on a particular research project.
  Advisors and mentors are typically less involved in the technical aspects of
  the research, including the creation and maintenance of artifacts, but are
  well positioned to promote and support primary researchers in their
  artifact-related efforts.

\item[\textbf{Reviewers:}] Potential artifact reviewers include Artifact
  Evaluation Commitee members, journal paper reviewers, and technical program
  committee members who may have access to the artifact.

\item[\textbf{Process Organizers:}]
  Individuals in this role are reponsible for the design, organization,
  and implementation of artifact review, and include \ac{aec} chairs,
  conference chairs, and journal editors.
  These individuals have the ability to improve the effectiveness and
  outcomes of the review process, and, by extension, raise the standard of
  artifacts within the community.

\item[\textbf{Community Leaders:}]
  Community leaders include steering committees, journal editors, professional
  organizations,
  reappointment and promotion committees, hiring committees, and funding agencies.
  These entities collectively hold significant influence over the
  research community and, as a result, interventions at this level are capable of achieving wide-reaching, systemic change.
\end{description}

\subsection{Threats to Validity}
\label{sec:methodology:threats}

As with all work, it is possible that we could have inadvertently biased our results due to our methodology.  We examine our threats to validity, and organize them into the following areas: replicability, construct, internal, and external.

\emph{\textbf{Replicability} Can others replicate our results?}
Qualitative studies, in general, can be difficult to replicate.
We have made as much of our materials available as possible, while still
preserving the confidentially of the survey respondents.
With our accompanying artifact, we have published
our publication survey dataset,
the design materials (i.e., questionnaire, consent forms, recruitment
email) and deidentified quantitative responses for our author survey,
and a Jupyter~\citep{jupyter} notebook for reproducing the figures presented in this paper.

\emph{\textbf{Construct} Are we asking the right questions?}
One potential threat to this work is how we measured artifacts.  We consider a
paper that has a link to accompanying work to have an artifact. However, it
could be the case that papers could have artifacts that are not documented in
the paper, which could potentially be found only by searching, or contacting
the authors of the paper. We did not contact authors for practical reasons, but
we acknowledge that there could exist artifacts that were not included in our
analysis.

As in every study, the construction of the study can impact the results.
To achieve the best results possible, we used a
mixed-methods parallel-convergent design incorporating three components:
(1) analysis of quantitative data describing artifact publication and evaluation behavior;
(2) analysis of survey data describing authors' perceptions,
needs, and experiences;
and (3) a subsequent integrated analysis.
This design was selected in order to obtain different but complementary data on
the subject, and to synthesise results for a more complete understanding of
artifact sharing~\citep{morse1991qualitative}.

\emph{\textbf{Internal} Did we skew the accuracy of our results with how we collected and analyzed information?}
Surveys can be affected by intentional and unintentional bias, both from the survey respondents and from the researchers. To mitigate this concern, we asked respondents to report their own experiences, feelings, values, and needs. We analyzed the qualitative components of our study
collaboratively using a concept-orientated approach.
We carefully documented
our process throughout the phases of data collection and analysis.  To ensure
the validity of our study, we used a purposeful sampling approach, triangulation of
researchers, and triangulation of analyses \citep{kitto2008quality}.

The reliability of qualitative and mixed-methods research is determined by the
rigor and consistency with which the methodological processes are applied. We collected data through multiple sources and interpreted our results by using multiple conceptual frames \citep{kitto2008quality}.
We enhanced the reliability of our process and results
through constant comparison, comprehensive data use, and use of tables, as
proposed by \cite{silverman2008doing}.

In this work, we define an artifact as being available if the link is still
alive.  It could possibly be the case that what is there is completely
unusable, or perhaps the artifact page links to a download that is no longer
available.  However, evaluating the quality of artifacts is beyond the scope of
this work, and is something that we leave to other researchers.

\emph{\textbf{External} Do our results generalize?}
Because of the nature of surveys, we cannot generalize our results beyond our
survey population.
Perhaps if we had chosen a different population, we would have had different answers.
To mitigate this, we sampled authors from
several of the top software engineering conferences (\ac{icse}, \ac{fse}, \ac{ase}), as well
as a leading software engineering journal (\ac{emse}).

Our study was
designed to describe the values, experiences, and needs of the software
engineering research community: It does not return results which can be
generalized beyond this particular domain.

\section{Results}
\label{sec:results}

In this section, we present the results of our primary analysis, organized by each
research question. \Cref{sec:results:rq1} examines how the software engineering
research community perceives and values artifacts, and the reasons for
which artifacts are used and shared.
\Cref{sec:results:rq2} subsequently identifies the challenges that are faced by
artifact creators, users, and reviewers.

\subsection{RQ1: Does the software engineering research community perceive inherent value in artifacts?}
\label{sec:results:rq1}

\begin{figure}[t]
\centering
\includegraphics[width=0.7\textwidth]{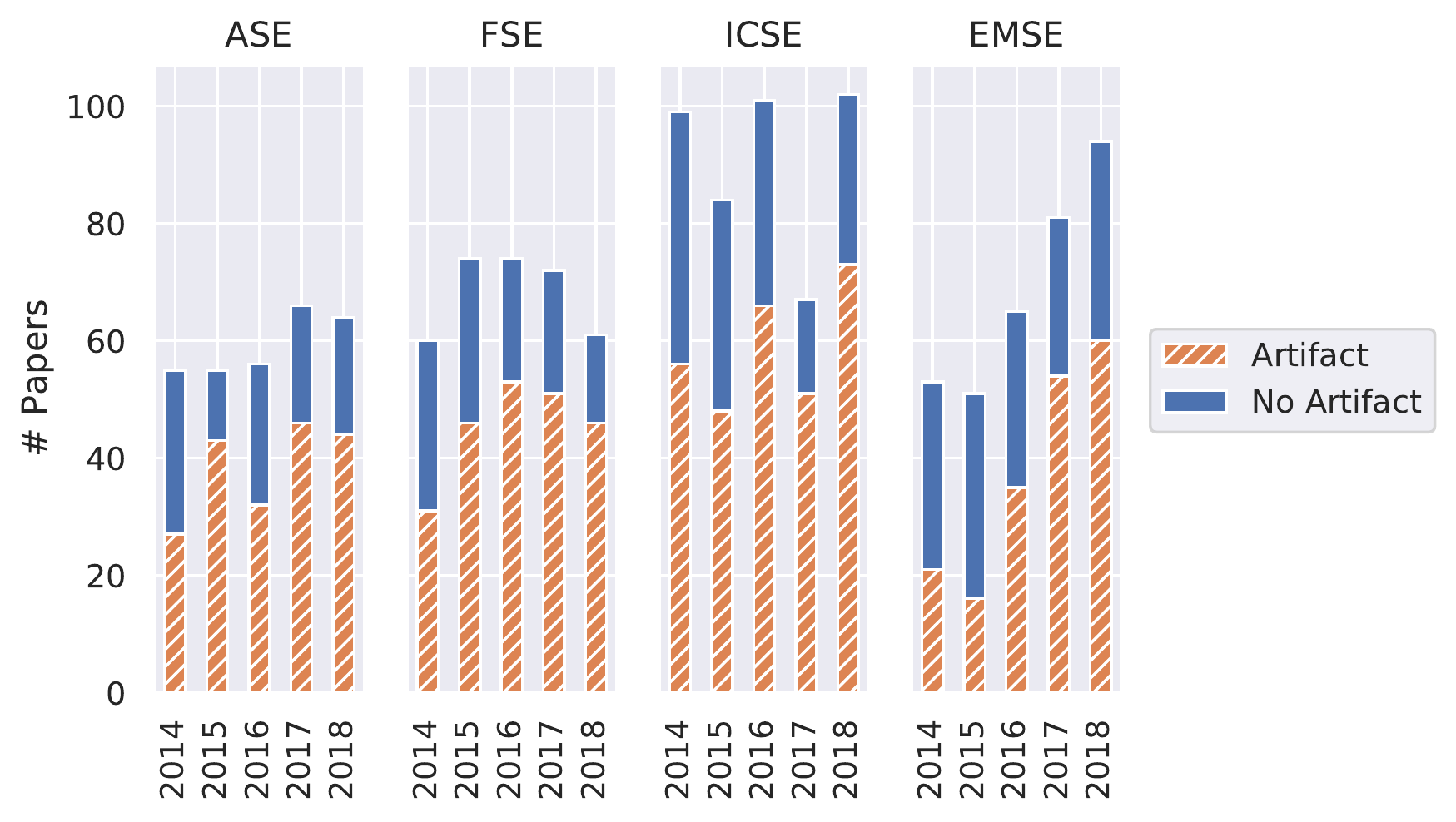}%
  \caption{A breakdown of all technical papers published at \ac{ase}, \ac{fse}, \ac{icse},
  and \ac{emse}, showing the fraction of papers that do contain an artifact
  (orange, dashed) vs. those that do not (blue), between 2014 and 2018, inclusive.}
\label{fig:artifact-prevalence}
\end{figure}

From the results of our publication survey, illustrated in \Cref{fig:artifact-prevalence},
we find that almost two thirds
(62.69\%) of papers published at \ac{ase}, \ac{emse}, \ac{fse}, and \ac{icse} between 2014 and 2018
(inclusive) contain an artifact.
Furthermore, across all venues that we studied, the proportion of papers
containing an artifact grew from 50.56\% in 2014 to
69.47\% in 2018,
indicating an increasing prevalence of artifacts.

When we asked survey respondents whether they value artifacts and what they
think artifacts should have, do, or be, it became clear that the software
engineering research community does indeed perceive inherent value in artifacts.
While participants recognize that not every paper needs an artifact,
numerous participants expressed positive sentiment around the sharing of
artifacts, and believe that sharing should be encouraged and rewarded by
the community.
P89 expresses that they are \enquote{[...] a big fan of sharing!}
and
P20 says that \enquote{[s]haring is essential [...].}
In the words of P82,
\enquote{[...] artifacts are moving toward becoming a necessary
and important part of publishing a research paper. It allows for meaningful
analysis and evaluation of the approach. Also, replication of studies are
extremely difficult (and often unreliable) without the appropriate
information contained within artifacts.}
Indeed, P34 highlights that
\enquote{[e]mpirical research may need more data artifacts.}

\begin{table}[t]
\themeheader
\themerow{Supporting evidence}{
Artifacts can be used to provide supporting evidence for the claims in a
research paper, and to improve the confidence of readers and reviewers.}
{62}
{I think publishing artifacts is good choice.
It would increase readers' confidence on how well proposed approaches are validated.
I do not fully believe results in some papers, which even are published on
top conferences such as \ac{icse} and \ac{fse}.}

\themerow{Community norms}{
Some researchers see the creation and sharing of artifacts as a
commendable pursuit
that benefits the community.
}{99}{It's good practice}

\themerow{Facilitate reuse}{
Researchers may share tools and datasets with their papers that can be reused
and extended by others.
}
{130}
{To provide a dataset}

\themerow{Improving understanding}
{Artifacts may be used to provide additional information about a study,
such as technical details that are not suitable for a paper,
or raw and preprocessed results.}
{136}
{The paper was about how we created a tool. That tool is now publicly available (open source)}
\bottomrule
\end{tabularx}
\caption{A high-level summary of the motivations for sharing artifacts.}
\label{tab:motivations:sharing}
\end{table}

To better understand the motivations for sharing artifacts, we asked participants for
the reasons they had shared an artifact with their most recently published
paper that contained an artifact.
Table~\ref{tab:motivations:sharing} provides a high-level summary of the motivations
for sharing artifacts based on a thematic analysis of responses.

\begin{insights}[left=0.5mm]{Insights for RQ1}
\begin{itemize}[label=\textbullet]
\item An increasing majority of research papers contain artifacts
\item Artifacts are perceived as inherently valuable by the research community
\item Authors create and share artifacts to support their claims,
allow others to better understand their work, and to facilitate
further research
\item Creating and sharing of artifacts is an established community norm
\end{itemize}
\end{insights}

\subsection{RQ2: What are the challenges of creating, sharing, using, and reviewing artifacts?}
\label{sec:results:rq2}

\begin{table}[htp]
\renewcommand\arraystretch{1.6}
\setlength{\aboverulesep}{0pt}
\setlength{\belowrulesep}{0pt}
\rowcolors{2}{white}{lightgray!30}
\begin{tabularx}{\linewidth}{p{0.005\textwidth}p{0.45\textwidth}X}
\toprule
& \textbf{Title and description} & \textbf{Representative quote} \\[.5ex]
\midrule
\gthemerow{\gcreator}{C1: Not worth it}
  {The time required to prepare an artifact could be better spent on other high-value
  activities (e.g., paper writing).
  Additionally, the risks of sharing an artifact may outweigh the rewards of doing so.}
  {137}
  {Getting the artifacts (raw data, diagrams, tools) into
  publishable form requires extra time (e.g. for cleaning, documentation,
  designing a web page), which I typically don't have while preparing the main
  publication.  Putting my time into improving the main publication has a clear
  benefit, whereas the benefit of releasing the artifacts is hard to quantify.}

\gthemerowx{\gcreator\newline\guser\newline\greviewer}{C2: Portability}
  {Artifact may not build or run as intended on other machines
  due to missing information, poor packaging, a reliance on outdated and
  hard-to-obtain dependencies, and bugs in the source code.}
  {\presponse{4}{The code requires refactoring to be executed independently
  from the experiment environment.}}

\gthemerow{\gcreator\newline\guser}{C3: Maintenance}
  {The creator of the artifact may have moved on (e.g., a graduating Ph.D. student),
  or is no longer interested in or able to maintain the artifact.}
  {107}
  {Some [artifacts] are no longer maintained, which is understandable.}

\gthemerow{\gcreator\newline\guser\newline\greviewer}{C4: Tacit knowledge}
  {Poor documentation or a lack of documentation, coupled with
  incomprehensible code, make it difficult to understand how the artifact
  works, and how it can be extended and reused.}
  {48}
  {Huge R files that are not comprehensible at all.}

\gthemerow{\guser}{C5: Artifact does not fit purpose}
  {The artifact does not match the claims in the paper, or is difficult to use
  as intended.}
  {14}
  {In four cases this year, I re-implemented the code and even
  re-collected the data, because the data was pre-processed and not raw data.}

\gthemerow{\gcreator\newline\greviewer}{C6: Lack of standards and guidelines}
  {Creators and reviewers are unsure of how artifacts should be packaged, and what standards
  those artifacts should meet.}
  {61}
  {It is quite difficult to create an artifact for evaluation,
  particularly when the standard of acceptance is not clear.}

\gthemerow{\gcreator\newline\guser}{C7: Hosting}
  {It can be challenging to find a host for the artifact that ensures long-term
  archival, supports large file sizes, and is affordable or free. Many
  artifacts are no longer accessible.}
  {8}
  {Some artifacts are just too big (like large datasets) and
  downloading and using them is impossible.}

\gthemerow{\gcreator}{C8: Double-blind review}
  {Submitting an artifact as part of a paper for double-blind review adds
  additional work and difficulty to the process of sharing artifacts.}
  {124}
  {Sharing the artifacts was too much of a hassle due to
  double blinded review.}

\gthemerow{\gcreator\newline\guser\newline\greviewer}{C9: External constraints}
  {Certain artifacts are difficult to share due to reasons beyond the control of
  their creators.}
  {16}
  {I typically am unable to include actual data for
  privacy/legal reasons due to the fact I work at a company.}

\gthemerow{\greviewer}{C10: Lack of reviewer incentives}
  {Reviewing artifacts is seen as a time-consuming activity with little reward}%
  {37}
  {We usually just make students do AEC. This is good in some ways because it gets students involved in the reviewing process. But it also reinforces the idea that artifacts are not a primary thing and that authors should really focus on the paper.}

\gthemerow{\greviewer}{C11: Technical obstacles}
  {Inherent challenges that make it difficult to evaluate artifacts with limited time and resources.}
  {50}
  {It is hard to decide what to do for artifacts that by nature take more
  time than one can devote to evaluating a single artifact ($\sim$ one day).}

\gthemerow{\greviewer}{C12: Limited communication}
  {A lack of open communication between creators and reviewers makes
  it hard to resolve inevitable technical issues and leads to
  frustration for both parties.}
  {142}
  {Sometimes \acp{aec} adopt a single review (as against multiple
   rounds of contacting authors). This can be really frustrating especially because it
   takes a considerable amount of time to build an artifact and a minor glitch in the
   installation instructions should not be grounds for rejection}
\bottomrule
\end{tabularx}
\caption{The high-level challenges of artifact creation, sharing,
  use, and review that are experienced by creators (\gcreator),
  users (\guser), and reviewers (\greviewer).}
\label{tab:challenges}
\end{table}

\Cref{tab:challenges} provides an overview of the high-level challenges that affect the
creation, sharing, use, and review of artifacts, that emerged from our thematic
analysis of survey responses.
In the following sections, we discuss how these challenges affect artifact
creators (\cref{sec:rq2:creators}), users (\cref{sec:rq2:users}), and
reviewers (\cref{sec:rq2:reviewers}).

\subsubsection{RQ2.1: What challenges are faced by artifact creators?}
\label{sec:rq2:creators}

Below, we describe the high-level challenges in \Cref{tab:challenges}
that affect artifact creators.

\paragraph{\textbf{C1: Not worth it}}

Although artifacts are seen as being beneficial to the community,
authors perceive that
creating and sharing artifacts can be difficult and time consuming,
producing artifacts does little to advance one's own career,
and that, ultimately, the time spent preparing artifacts
could be better spent on other higher-value research activites (e.g.,
paper writing).
That is, there is a high \emph{opportunity cost} associated with creating
and sharing artifacts.
This cost is at its highest when artifacts are submitted for evaluation:
As P1 puts it,
\enquote{the time required to submit to an artifact track is not rewarded enough. In other words, there is not enough gain in terms of publication quality to justify it.}
P23 sees the overheads of artifact review as a potential reason not to submit
artifacts at all:
\enquote{I've never been asked to do so, but it sounds like a lot of extra hassle and work.  This extra cost would make me re-think whether I wanted to submit artifacts at all.}

Authors also report a variety of risks that outweigh the potential
rewards of sharing artifacts.
The perceived risks include
that
poor code quality may harm their reputation,
a fear of being \enquote{scooped,}
a fear of mistakes in the analysis being discovered,
the burden of maintaining the artifact,
and the possibility that the artifact may be used by no one.
For example, when asked why they had not shared artifacts with a recent paper,
P87 said that they \enquote{didn't have enough time to publish it, [were] scared the analysis
is incorrect, [and] wanted to save it for the next publication.}

Authors may be reluctant to share artifacts that still have some
untapped publication value, or may wish to avoid being \enquote{scooped}
by others by providing their artifacts prematurely. As P120 puts it,
\enquote{sometimes, one needs to protect a Ph.D. students abilities to
produce sufficient results prior to making artifacts available to others.}

P61 chose not to share an artifact with a recent paper due to the belief that
there is a limited audience:
\enquote{There is only a small group of people working on the topic of
  the paper. Perhaps [no]body other [than] myself will keep working on the same topic
  and use the artifact.}
In contrast to this view, P68 highlights the importance of sharing
artifacts, particularly for emerging areas of research:
\enquote{Despite my complaining, moving towards a norm of making artifacts available is
incredibly valuable. I know of whole research areas that are dead because
the early work had no artifacts available, and the cost of re-implementing
their work just so you can move on to something novel is too high.%
}

Overall, despite the perception that the creation and sharing of artifacts is
a poor investment of time and effort, some researchers, such as P5, have seen
personal benefit from sharing artifacts with their research:

\begin{quote}
\enquote{%
I had some experience where some researchers were not in favor of
providing research artifacts, either because the effort/time investment was not
worth it, or because it may \enquote{take away} from their next paper.

On the other hand, in my own experience so far, providing these artifacts
(companion website, accompanying blog post, or a replication package) seems to
have been very beneficial for the papers and the dissemination of our results.%
}
\end{quote}

\paragraph{\textbf{C2: Portability}}
Anticipating and preparing for the possible environments in which an artifact
may be used is a considerable challenge, as P70 shares:
\enquote{The major challenge I face is in ensuring that [the] artifact runs in other
environments which is hard to verify beforehand.}

Retroactively addressing such concerns in existing artifacts may require extensive
refactoring, or the creation of a virtual machine or container image (e.g., Docker~\sqcitep{docker}).
In either case, the repackaged artifact should ideally be tested in a variety
of environments. P37 shares that,
\enquote{the challenge was to find a suitable way to share the
  artifacts so that they are accessible/runnable by everyone. Sometimes we have
  to try the artifacts on different environments to make sure they work.
  Alternatively we can provide VM or Docker images, which might have their own
  challenges (e.g., VM images becoming too big, etc). Also, time pressure does
  not allow documenting everything for running/using the artifacts, making it
  difficult to make sure everyone can use them eventually.}

\paragraph{\textbf{C3: Maintenance}}

Sharing an artifact not only requires an upfront investment of time and resources
to prepare the artifact, but also an ongoing investment to continually maintain
the artifact.
As bugs are uncovered, use cases evolve, and inevitable bit rot occurs,
there is a need to update the artifact.
When asked why they had chosen not to create
and share an artifact with a recent paper, P50 said that
\enquote{[t]ime is the main reason; it takes a lot of time to package [the] artifact
so that it can be useful to others,} before going on to point out that
\enquote{if students graduate, it is hard to find somebody else to prepare the artifact
and make it available online.}
Indeed, as students are often those responsible for the creation, sharing, and
maintenance of artifacts, it can be difficult to maintain those artifacts
once those students move on.

\paragraph{\textbf{C4: Tacit knowledge}}

The process of writing documentation
and refactoring code can be challenging and burdensome.
As P35 says of
creating and sharing artifacts,
\enquote{the bottleneck tends to be creating the required
  documentation (both usage instructions and code documentation).}
In the absence of concerns around the time and effort required to produce
documentation, tacit knowledge remains a major obstacle to artifact
creators~\citep{Shull02}. As P133 puts it,
\enquote{the main difficulty is to provide a highly automated way
  to utilize the artifact. A lot of knowledge is typically buried in developers' heads.}
The problem of tacit knowledge is compounded
by the quickly evolving, prototypical nature of most artifacts, as
described by P103:
\enquote{We usually develop prototype tools and usually as researchers
  we understand their restrictions and limitations to a certain level. For
  example, we have not tested them in all environments. It is very difficult
  to have a manual.}

The need to anticipate the various users apriori at the time of producing
the artifact can be a challenge, as P50 identifies
\enquote{there is always a question [of] what is the best way
  to package the artifact to be useful to others.
  Once you take a step forward, it is hard to go back and recreate everything.}

\paragraph{\textbf{C6: Lack of standards and guidelines}}

Creators identify a lack of clear standards and expectations
around the packaging, contents, and quality of artifacts as a difficulty when
deciding to share their artifacts for formal review:
As P61 says,
\enquote{it is quite difficult to create an artifact for evaluation,
  particularly when the standard of acceptance is not clear.}

In some cases, creators are unclear as to what is considered to be an
artifact for the purposes of formal review, and how to proceed in cases
where the work builds on top of an existing artifact.
For example, P34 says that it is \enquote{unclear how to handle artifacts that use
research data from public datasets / existing artifacts.
Should we create a new artifact, or point towards an existing one?}
Creators blame such ambiguity and unclear expectations on a lack of
sufficient guidance on the contents and packaging of artifacts within
the community. P104 shares:

\begin{quote}
\enquote{%
To me, I think one of the biggest challenges in creating artifacts
that accompany papers relates to the amount of work required going from
often messy research code to something that can be used with relative ease by
a larger body of interested individuals. This can often be a significant
amount of work, and I think the community lacks some guidelines related to what
actually needs to be included with software artifacts in order for them to be
as usable and reproducible as possible. For example, things like setup
instructions, contribution guides, sufficient documentation, etc, are often
overlooked.%
}
\end{quote}

\paragraph{\textbf{C7: Hosting}}

To facilitate sharing, creators must find a suitable place for long-term hosting
of the artifact that meets their various needs:
(1) that the artifact must be reliably available indefinitely,
(2) the host must be able to store large volumes of data,
and (3) the hosting service must be affordable, or, ideally, without cost
to researchers.
As P137 states, finding such a host can be difficult:
\enquote{I am unaware of (free) services that can guarantee long-term
accessibility to my artifacts.}
Likewise, P110 shares,
\enquote{It is usually hard to find a good host for them [artifacts] that is
\enquote{respectfu[l]} and will host them [artifacts] for long,
}
and P12 says, \enquote{one challenge is to share a large dataset that does not fit into a GitHub repo or Dropbox.}
Note that, while popular services such as
Google Drive,\footnote{\url{https://drive.google.com} [Date Accessed: March \nth{13} 2021]}
GitHub,\footnote{\url{https://github.com} [Date Accessed: March \nth{13} 2021]}
and Dropbox\footnote{\url{https://dropbox.com} [Accessed: 12 Mar 2021]}
provide a
free or inexpensive method of hosting artifacts, they are not
intended as an long-term archive, and are ill-suited to large datasets.

\paragraph{\textbf{C8: Double-blind review}}

Creators report that the double-blind process adds additional difficulties
to the process of sharing artifacts in two ways.
Firstly, artifacts must be carefully anonymized such that the identities of
their authors are not revealed. As P5 puts it,
\enquote{we need to create the artifacts in such a way that it won't violate
the double-blind review process.}
Secondly, the double-blind review process further complicates the difficulties
of finding a suitable host for the data by adding the requirement of
anonymity.
P155 says, \enquote{it takes me some time to find a suitable place to
anonymously provide artifacts for double-blind review.
I use Google Drive most, but it does not provide an option for anonymous
share.}
Despite the challenges, some creators report success in finding suitable
hosting services:
P19 says,
\enquote{before I discovered Zenodo, it was difficult to share artifacts[,]
  especially when the conferences had [a] double-blind policy.}

Ultimately, the additional effort needed to overcome these difficulties can
dissuade creators from sharing their artifacts at the time of paper submission.
For example, P124 says that \enquote{sharing the artifacts was too much of a hassle
due to double blinded review.}

\paragraph{\textbf{C9: External constraints}}

Circumstances beyond the control of creators may prevent or complicate
sharing artifacts with the general public.
Such circumstances may be ethical or legal in nature,
such as intellectual property restrictions and privacy concerns,
as well as more technical circumstances, such as the artifact belonging to a larger
ecosystem, making it difficult to share and reproduce.

Creators working in or collaborating with industry
reported being unable to share the associated artifacts
(e.g., code and data) of their research with the general public due to
\ac{ip} restrictions. 
For example, P140 says, \enquote{in our collaboration with an industry partner[,]
  sharing artifacts was not allowed for contractual reasons.}
Fear of liability was also given as a reason for being unable to share artifacts
by more than one creator, such as P41 who shares that
\enquote{the paper that we did not share artifacts for was based on data obtained from our industrial collaborator. We asked them if we could share this in anonymised form but they did not allow this, fearing that the data may still be misused (e.g. to start liability lawsuits).}

Creators also voiced their frustration at reviewers for a perceived lack
of understanding of such external circumstances. P22 shares
that they were
\enquote{once criticised for not sharing data for an industry track
  paper coauthored with practitioners. Artifact should not be used as
  ritualistic blanket argument to judge papers.}

Privacy concerns and the need to de-identify personal data were also given as
difficulties of sharing data artifacts, and, in some cases,
a reason for not sharing at all. When asked about the challenges of creating
artifacts, P44 cited:
\enquote{de-identifying survey results and worrying whether we were thorough enough.}
Indeed, beyond being an error-prone and time-consuming process, studies have
shown that sharing qualitative data risks re-identification even when steps
are taken to remove personal identifiers from the
data~\citep{narayanan2008robust,el2011systematic,ji2014structural}.
Given the inherent risks of sharing sensitive data, it is perhaps unsurprising
that participants reported not sharing data artifacts, such as P44, who states
\enquote{I don't share qualitative data, as a rule.
It seems too risky (re-identification) and would run counter to IRB conditions.}

\begin{insights}[left=0.5mm]{Insights for RQ2.1}
\begin{itemize}[label=\textbullet]
\item Artifact creation and sharing is perceived as a risky and poor
  investment of time and effort that produces little personal reward

\item Sharing certain artifacts necessitates maintenance, which is difficult to provide when
  students graduate and move elsewhere

\item A lack of community standards and incentives makes it difficult for
  creators to produce high-quality artifacts that can be used by others

\item Finding a hosting service
  that is free, capable of storing large volumes of data, and provides indefinite archival
  of artifacts can be challenging

\item Some artifacts are dangerous, difficult, or impossible to share due to privacy and \ac{ip}
  concerns, or belonging to a larger ecosystem
\end{itemize}
\end{insights}

\subsubsection{RQ2.2: What challenges are faced by artifact users?}
\label{sec:rq2:users}

\begin{table}[t!]
\centering
\renewcommand\arraystretch{1.6}
\setlength{\aboverulesep}{0pt}
\setlength{\belowrulesep}{0pt}
\rowcolors{2}{white}{lightgray!30}
\begin{tabular}{lp{0.75\textwidth}r}
\toprule
 & \textbf{Challenges when using artifacts produced by others} & \textbf{\#} \\
\midrule
C4 & A lack of clear instructions for using the artifact                             & 103 \\
C7 & The artifact was no longer accessible                                           & 81  \\
C2 C3 C4 & A lack of clear instructions for building the artifact                          & 79  \\
C2 & Couldn't build the artifact due to unavailable software, library, or compiler   & 69  \\
C2 & Couldn't run the artifact due to missing package, library, tool, or software    & 66  \\
C2 C5 & Had to modify the source code of the artifact                                   & 60  \\
C2 & Encountered a run-time error when using the artifact                            & 55  \\
C2 C5 & Had to re-implement the artifact                                                & 28  \\
   & I have never tried and failed to use an artifact                                & 10  \\
   & Other                                                                           & 10  \\
\bottomrule
\end{tabular}
\caption{A summary of 130 responses to the question:
  \enquote{What challenges have you faced when trying to use an artifact other than your
  own?} annotated with their frequency and the identifiers of their associated challenges.}
\label{tab:quant:use}
\end{table}

Below, we describe the high-level challenges in \Cref{tab:challenges}
that affect artifact users.
Quantitative results from Q11,
\enquote{What challenges have you faced when trying to use an artifact other than your
own?}
presented in \Cref{tab:quant:use}, are discussed within the context
of the high-level challenges.

\paragraph{\textbf{C2: Portability}}
\label{sec:users:portability}

As P9 highlights, artifacts may suffer from a variety of portability issues
that prevent them from being reused in other environments (e.g., hardware,
OS, and software differences):
\enquote{Many artifacts have had hardwired (and undocumented) dependencies on
particular files and folders such as the code author's home directory.
Some require the user to be \enquote{root}.}
From our quantitative results, shown in \Cref{tab:quant:use}, we observe
that many artifact users have experienced portability-related
challenges. 
79 users complained about a lack of clear instructions for building an
artifact, 69 users were unable to build an artifact due to
dependencies that are no longer available (e.g., libraries, compilers,
tools, etc.), and 66 users reported being unable to run an artifact
due to missing dependencies.
In the event where users were able to build and run an artifact,
55 users encountered a run-time error.
60 users reported that they had to modify the source code of an
artifact, and, in the most extreme cases, 28 users shared that they
had to reimplement an artifact entirely.

\paragraph{\textbf{C3: Maintenance}}

A lack of maintenance can impede or prevent users from being able to use an
artifact.
For example, P137 recalls a challenge in using an artifact where the \enquote{[r]equired
libraries were not exactly unavailable, but hard to find.  The artifact was no
longer actively maintained and used old version of libraries, which become
harder and harder to get to work.}

Despite the difficulties of dealing with unmaintained artifacts,
users are sympathetic of the challenges that face creators.
P105 acknowledges that \enquote{some [artifacts] are no longer maintained, which is understandable,}
and P102 stresses that \enquote{[t]here is more of a need to recognize that more funding is
needed to focus on artifact maintenance.}
P20 identifies that, while maintaining artifacts is important,
few researchers have the resources required to do so:
\enquote{Sharing is essential, and it should be more valued by the community.
In this \enquote{publish or perish} system, few care is given to the paper's
artifacts as the focus is, indeed, on the next paper.
IMO, only big groups, or only people who have [money] to spend for artifact
maintenance, are doing a good job, which is clearly far from being the
ideal situation.}

\paragraph{\textbf{C4: Tacit knowledge}}
A large number of users complain
about difficulties in understanding, building, and using artifacts created
by others due to a lack of documentation and poor code
quality, as shown in \Cref{tab:quant:use}.
For example,
P70 shares that they \enquote{often find the necessary documentation missing[,] which makes it hard to comprehend and run the artifact,}
and P33 says that \enquote{some authors think that just adding a link to the raw material is sufficient. Guidance is needed to understand
artifacts, and this takes time.}
The magnitude of this challenge is highlighted by our quantitative results,
presented in \Cref{tab:quant:use},
which show that
\enquote{a lack of clear instructions for using the artifact}
and
\enquote{a lack of clear instructions for building the artifact}
are the first and third-most common challenges experienced when attempting
to use an artifact produced by someone else.

\paragraph{\textbf{C5: Artifact does not fit purpose}}

In some cases, artifacts may be difficult for others to use for the
purpose stated in their associated research papers.
P63 states that \enquote{they [artifacts] don't really serve the
purpose stated in papers,}
and P2 points out that \enquote{most artifacts work for very limited use cases.}
Using an artifact for its intended purpose may involve extensive changes
on the part of the user:
62 participants report having modified the source code of an artifact,
and 28 participants had to reimplement the artifact,
as shown in \Cref{tab:quant:use}.

\paragraph{\textbf{C7: Hosting}}

When asked what challenges they had faced when attempting to use an artifact,
82 users stated that being unable to access an artifact, making it the
second-most commonly reported challenge (\Cref{tab:quant:use}).

\begin{figure}[t]
\centering
\includegraphics[width=0.6\textwidth]{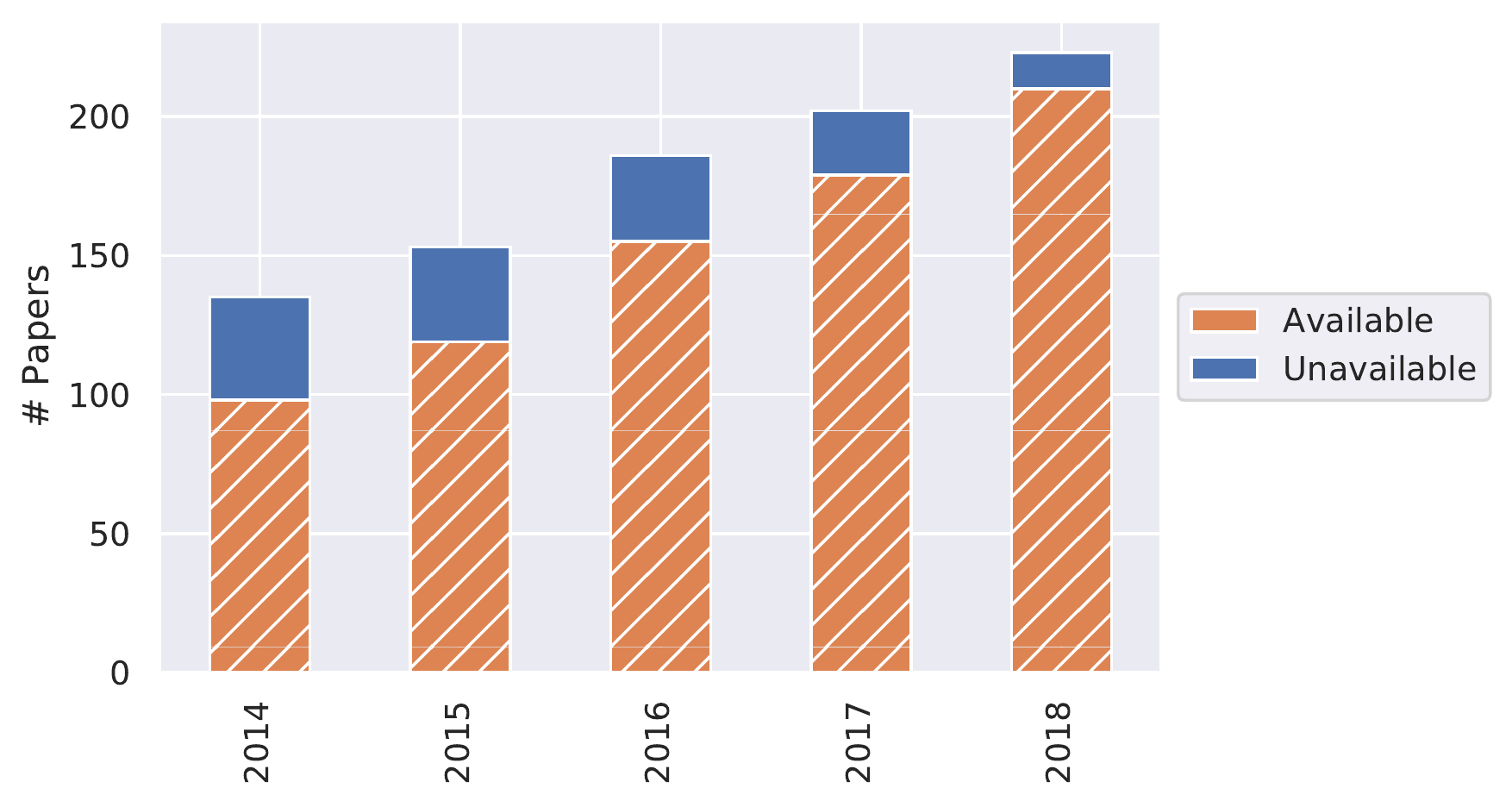}
\caption{
  An overview of the availability of all artifacts within our dataset
  over time, determined by manually checking whether the artifact URL
  could be accessed at the time of inspection (between January \nth{29} and February \nth{6}, 2019).%
}
\label{fig:availability}
\end{figure}

From the results of our publication survey, given in \Cref{fig:availability},
we find that approximately 14\% of all artifacts within our dataset are inaccessible
via the URL given in the corresponding paper. As one may expect,
older artifacts are more likely to be unavailable (26.47\% in 2014),
but we also observe that several more-recent artifacts (5.43\% in 2018)
are also unavailable.

\paragraph{\textbf{C9: External constraints}}

Licensing restrictions can create difficulties for those attempting to build upon and extend artifacts.
For example, P35 shares that
\enquote{to use and extend the artifact, I had to recover the
  source code using a decompiler. Afterwards I couldn't make available a
  replication package for the extended artifact due to licensing issues.}
Likewise, P9 identifies that some artifacts
\enquote{require a paid commercial license [for the] auxiliary tools necessary to use
[the artifact].}

\begin{insights}[left=0.5mm]{Insights for RQ2.2}
\begin{itemize}[label=\textbullet]
\item Artifacts may become unavailable and unmaintained over time
\item Missing documentation, non-portable code, and licensing restrictions make it difficult to
  reuse and extend artifacts
\item Users often need to modify artifacts to fit their needs and expectations
\item These challenges are exacerbated by the lack of community standards for packaging
  and sharing artifacts
\end{itemize}
\end{insights}

\subsubsection{RQ2.3: What challenges are faced by artifact reviewers?}
\label{sec:rq2:reviewers}

\begin{figure}[t!]
\centering
\includegraphics[width=0.7\textwidth]{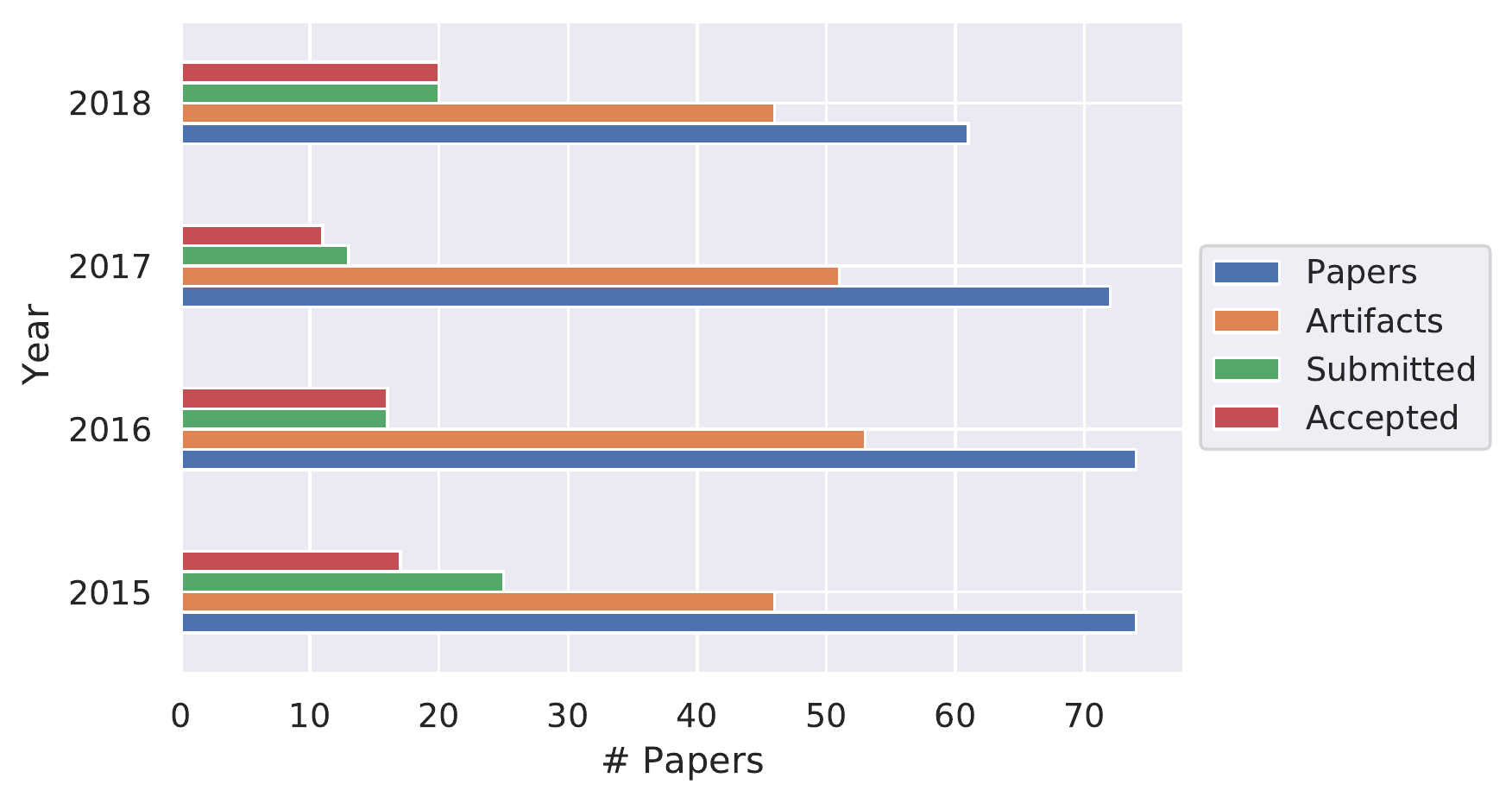}
\caption{A summary of the papers accepted at \ac{fse} between 2015 and 2018 that
  were submitted to and accepted by the \ac{aec} for the conference
  where
  \textbf{Papers} is the number of papers accepted at the conference,
  \textbf{Artifacts} is the subset of those papers we deem to contain an artifact,
  \textbf{Submitted} is the subset of papers that contain artifacts that were
  submitted to the \ac{aec},
  and \textbf{Accepted} is the subset of papers that were accepted by the \ac{aec}.
  Note that neither \ac{icse} nor \ac{ase} conducted artifact evaluation between
  2015 and 2018.}
\label{fig:aec-acceptance}
\end{figure}

As the predominant mechanism for assessing and upholding the quality and claims
of software artifacts, \acp{aec}
are well positioned to mitigate the downstream issues faced by those using
artifacts.
However, despite our observation that almost two thirds of
research papers have an associated artifact (\cref{sec:results:rq1}), we
find that relatively few papers with artifacts are submitted for
evaluation when it is possible to do so.
Of the 281 research papers accepted at conferences that had an \ac{aec}
(i.e., \ac{fse} 2015--2018),
of which we deem 196 to contain artifacts via our analysis (69.75\%),
74 papers were submitted to the \ac{aec} (26.33\%),
and 64 were subsequently accepted, representing a markedly high acceptance rate of
86.49\%, shown in \Cref{fig:aec-acceptance}.

Below, we explore the low participation rate for artifact evaluation
by describing the high-level challenges that affect artifact reviewing.
We find that reviewers face many of the same challenges that are encountered
when using artifacts (e.g., problems stemming from a lack of portability and
documentation). To our surprise, we find that reviewers also face challenges
similar to those faced by artifact creators (e.g., inadequate guidelines and
incentives).
It should be noted that some reviewers reported being happy with the current
form of \acp{aec}:
When asked how the artifact evaluation process could be improved,
P157 stated \enquote{[i]t's pretty good I think,}
and P26 said that \enquote{[t]he current process is quite good.}

\paragraph{\textbf{C2: Portability}}

Reviewers experience the same portability issues that are experienced by
users, and highlight the process of installing artifacts and their dependencies as a
significant challenge during review.
P121 shares that
\enquote{[i]nstalling software and dependencies was the biggest challenge as a reviewer,}
P35 complains of an
\enquote{[o]verly complicated process for setting up the artifact and its environment,}
and P142 cites the
\enquote{requirement to install complex and invasive program/system libraries}
as a challenge when reviewing.
Artifacts are often only tested in the environment
in which they were produced, and, in some cases, may  be hardcoded to
that particular environment (e.g., the author's machine).
P107 recalls dealing with
\enquote{complicated instructions for installing and executing
  artifacts, that are platform depend[e]nt, and have not been properly tested on
  settings different than the authors' machines.}

Virtual machines (VMs) can be used to mitigate these portability challenges and
to avoid the security risks associated with executing untrusted
code on the reviewer's host machine.
However, VMs are not without their own problems.
P50 highlights that
\enquote{using virtual machines does not solve all problems
  (because running across different OSes leads to problems and sizes are huge),}
and P64 shares that
\enquote{some \ac{aec} folks try to evaluate the performance claims
  of the paper, but the software is typically in the form of a VM.
  Being in a VM, it is not really possible to accurately assess performance,
  nor should the authors be penalised if the reviewer cannot assess the
  performance.}
Containers may be used in lieu of virtual machines to address
performance and size concerns, but as P18 identifies:
\enquote{artifacts, esp. software artifacts degrade over time.
Something that compiles today is probably not to compile in five years, because
you do not have the same system. And (surprise!) having containers does not
always solve this problem.}

\paragraph{\textbf{C4: Tacit knowledge}}

A lack of adequate documentation to understand, install, and use an artifact
can make reviewing difficult or impossible.
For example,
P6 complains of \enquote{poor documentation from authors describing which claims in the paper
are supported by the artifact,}
and P98 recalls a case where they
\enquote{could not reproduce the paper analysis relying solely on the artifact
instructions, [and] had to read the paper to learn what was missing.}
Similarly, P60 shares that \enquote{even though that happened only once, I had to replicate a study
in which the replication package seemed to be complete. However, when I
followed the instructions, I couldn't find the same results found by the
authors. Unfortunately, the package was missing an instruction.}

Some reviewers suggest that automation, in the form of scripts that can replicate
the results of an experiment, may be used to mitigate the tacit knowledge problem.
However, other reviewers stress that these solutions are not a panacea.
P74 points out that, in the event where an artifact does not meet expectations,
\enquote{[i]t is often difficult to know whether the submitted device
  is faulty, or if [you don't know how to] use it.}
In other cases, an automated script may essentially behave as a blackbox that
produces the intended results, but provides the reviewer with no insight into
whether the process used to arrive at those results is sound.
As P25 identifies, a desire for automation may also prevent methodological errors from being
noticed:
\enquote{Artifact evaluation committees -- by asking for replicable scripts -- are likely to replicate the same mistake (if any) the original authors made.}
And while automation may make it easier to replicate the results of an experiment,
documentation is still needed to conduct a thorough evaluation of an artifact.
As P52 highlights,
\enquote{[i]t was easy to reproduce the experiments, but difficult to understand
  where should I add more case studies.}

\paragraph{\textbf{C6: Lack of standards and guidelines}}

Reviewers complain about ambiguous or missing guidelines for
reviewing artifacts, and a lack of criteria for what is considered to be
a good artifact.
For example, P141 says that the
\enquote{artifact evaluation scheme [is] difficult to understand,}
and P74 shares that \enquote{evaluation criteria are not always easy to apply.}
Likewise, artifact badging schemes may also lack sufficiently clear
and detailed criteria:
P49 complains that the
\enquote{description of badge requirement[s] on ACM website is very
  confusing, both for the reviewer and authors of the artifact,}
and P12 says
\enquote{sometimes it is hard to decide which badge to give since
    the information in the ACM Artifact Evaluation and Badging Guideline is too general.}

The purpose of the artifact review process itself is often unclear, and
reviewers may hold different views.
Artifact review may simply involve running a set of automated scripts to replicate
a result, or it may involve a more thorough investigation of the assumptions,
limitations, and generality of the artifact.
P6 points out that:

\begin{quote}
\enquote{%
[There are] [u]nclear expectations as to how thorough the reviewing process
should be.  Some \ac{aec} members uncritically run the scripts provided by the
authors. Others perform a deep-dive into the code and occasionally uncover
severe problems with the artifact. These deep dives are extremely
time-intensive, however, and often require multiple days of work. But if the
\ac{aec} doesn't do it, will anybody ever? Probably not.%
}
\end{quote}

Several reviewers, such as P32, believe that the goal of artifact review
\enquote{should be to see that each artifact gets accepted (via shepherding).}
But, as P6 notes, this can create tensions when artifacts do not meet the expectations
set out in the paper:
\enquote{Unclear relationship between the artifact review process and the paper shepherding process.
  An improperly implemented benchmark should be grounds for paper rejection.}
To resolve this tension, P11 suggests that artifacts
\enquote{should be evaluated as part of paper if [they are] the main result.}
At one extreme, P124 believes that
\enquote{[p]ublishing artifacts should be a precondition (if there are no legal
  etc. reasons to not publish them) to get papers accepted.}
Similarly, P146 shares that
\enquote{[a]rtifacts as replication packages should be mandatory to get a paper published in journal and conferences.}
At the opposite extreme, other reviewers, such as P143, believe that the
artifact review process should be faster and more lightweight means of
\enquote{rubber stamping} artifacts:
\enquote{It shou[l]d be less [of] a burd[e]n to get an artifact \enquote{stamp}.
Often, many papers provide links to repositories and so on.
Why could this not be automatically evaluated by a committee as soon as a paper gets accepted?}

\paragraph{\textbf{C9: External constraints}}

Licensing issues can make it difficult or impossible to review certain artifacts
as reviewers may not possess the necessary licenses to install and use the
artifact.
P65 highlights that there is
\enquote{no obvious pathway for what to do if the software artifact requires a
proprietary component (Windows, a commercial IDE). Some legal clarity around
this would be useful.}

\paragraph{\textbf{C10: Lack of reviewer incentives}}

Dealing with technical obstacles and artifacts that are unsuitable for
sharing (e.g., artifacts with missing documentation) requires
significant time and effort on behalf of the reviewer.
Reviewers highlight that these efforts are not recognized and rewarded to the
same extent as paper reviewing.
P115 states that \enquote{[t]he credit for AE reviewers is low,}
and P64 shares, \enquote{we usually just make students do AEC. This is good in some ways
because it gets students involved in the reviewing process. But it also
reinforces the idea that artifacts are not a primary thing and that authors
should really focus on the paper.}
A lack of incentives and sufficient recognition may lead to inconsistent
and low-quality reviews due to reviewers that are not fully invested in the process.

\paragraph{\textbf{C11: Technical obstacles}}

A variety of obstacles can complicate the process of
reviewing artifacts.
Several reviewers identify long execution times and a lack of adequate compute resources
as complicating factors when reviewing.
P50 shares,
\enquote{it is hard to decide what to do for artifacts that by
    nature take more time than one can devote to evaluating a single artifact ($\sim$ one day).}
Artifacts that require significant disk space, such as virtual machine images, can
also pose a number of challenges when obtaining and reviewing artifacts.
As P52 highlights,
\enquote{the authors provided a large VM that didn't fit in my disk.
  So I have to uninstall some things on my computer.}
Combined with slow download speeds, large file sizes can also make it difficult
to obtain artifacts.

In some cases, fundamental technical obstacles make it practically infeasible
to replicate an experiment via an artifact:
In the words of P64,
\enquote{an interesting tension is that the reviewers do not
  often have the capabilities to actually run the artifact. For example,
  many cloud-based contributions, fuzzers, automated program repair,
  require significant hardware to actually do anything useful.}

\paragraph{\textbf{C12: Limited communication}}

The challenges of reviewing are often exacerbated by the lack of open communication
channel between creators and reviewers.
P72 shares that
\enquote{sometimes \acp{aec} adopt a single review (as against multiple rounds of
contacting authors).
This can be really frustrating especially because it takes a considerable
amount of time to build an artifact and a minor glitch in the installation
instructions should not be grounds for rejection.}
Indeed, reviewers highlight the importance of communication.
P133 says that it is
\enquote{understandable that an artifact from a researcher is not
  perfect. An important thing is that the owners of the artifact are willing
  to help the users resolve outstanding issues,}
and P52 points out that \enquote{usually, people answer quickly when one report
a bug in an artifact that you are using.}

However, in cases where such communication is permitted, issues may still persist
due to unresponsive authors. For example, P25 shares that they
\enquote{found that there's a trade-off of putting weight on authors' and reviewers'
shoulders: Ideally, the reviewers should take into account the authors'
attempts in helping them get the tools to run.
But with unresponsive authors and in lack of a clearly defined protocol, the reviewers might have to wait several
days without knowing if there will eventually be a solution attempt -- which,
if successful, is then followed by half a day of performing the actual review.}

\begin{insights}[left=0.5mm]{Insights for RQ2.3}
\begin{itemize}[label=\textbullet]
\item Few papers (26\%) that contain artifacts are submitted for evaluation
\item Most papers (86\%) that are submitted for evaluation are accepted
\item There is a range of views on the purpose and extent of artifact review
\item Certain artifacts are difficult to assess due to time and resource limitations,
  and licensing issues

\item Artifact reviewing is perceived to be less rewarding than paper reviewing

\item A lack of open communication between creators and reviewers, coupled with
  unclear standards and expectations for packaging and assessing artifacts,
  leads to confusion and frustration between creators and reviewers
\end{itemize}
\end{insights}

\section{Secondary Analysis}
\label{sec:secondary}

In this section, we develop a more complete understanding of artifacts by considering the results of our parallel studies in concert and exploring relationships
among the themes and components we discover.
Specifically, we use \ac{doi} to understand the factors that influence the creation, sharing, use, and review of artifacts (\cref{sec:interpretation:innovations}),
the communication channels through which artifacts are understood and shared (\cref{sec:interpretation:communication}),
how artifacts evolve over time (\cref{sec:interpretation:time}),
and the role of artifacts within the social system of the software engineering
research community.
When we approach this landscape of actors, experiences, and values
holistically, it becomes clear that many of the challenges identified in our
primary analysis are related to each other, and that challenges impact
people differently depending on whether they are creating, sharing, using,
or reviewing artifacts.

\subsection{Innovations}
\label{sec:interpretation:innovations}

Below, we describe the qualities of artifacts that influence their diffusion
in terms of the characteristics of innovations within \ac{doi} theory:
\emph{compatibility}, \emph{trialability}, \emph{complexity},
and \emph{relative advantage}.

\textbf{Compatibility:}
Artifacts should be familiar enough to fit within the existing needs and
expectations of their intended users.
Users desire artifacts that achieve this by adhering to
existing packaging, interface, and use case conventions.

\textbf{Trialability:}
Artifacts should clearly advertise their contents, purpose, and claims to
allow potential users to quickly determine whether an artifact suits their
individual needs. Respondents identify that ideal artifacts provide a clear directory
structure that is described as part of the README, along with a brief
summary of the intended use cases, claims, and limitations of the artifact.

\textbf{Complexity:}
Artifacts should be reasonably easy for their intended audience to understand and use.
For tool artifacts, this includes providing ample documentation
and examples, including tests, and ensuring the artifacts are self-contained
and quick to download and run.
For data artifacts, this includes providing both raw and preprocessed data;
describing the format of the data, how it was collected,
identifying any assumptions that were made, and providing sample queries that
demonstrate how the artifact can be used.

\textbf{Relative Advantage:}
Artifacts should offer tangible benefits to their users over choosing not
to use them, whether that be by providing a deeper understanding of an
associated study than is available in the paper, or by providing a reusable
tool or dataset and thereby avoiding the need to create one from scratch.

\subsection{Communication Channels}
\label{sec:interpretation:communication}

\begin{figure}[t]
\includegraphics[width=\textwidth]{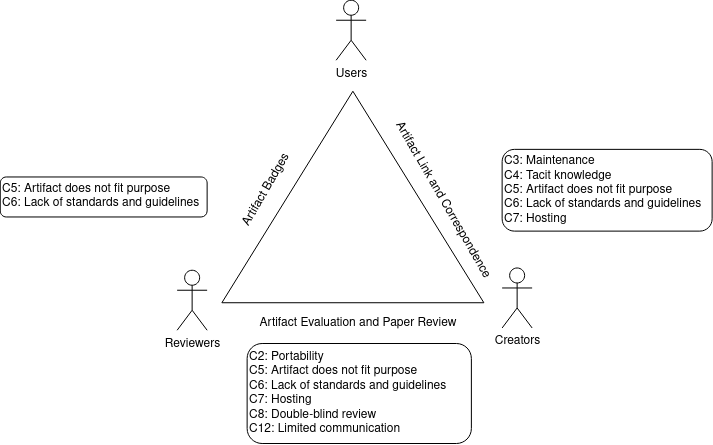}
\caption{An overview of the communication channels that exist between artifact
  creators, users, and reviewers, annotated with the high-level challenges that
  either affect or are affected by those communication channels.}
\label{fig:doi:communications}
\end{figure}

\Cref{fig:doi:communications} provides an overview of the communication channels that
exist between artifact creators, users, and reviewers.
Artifact evaluation and paper review serve as a channel between creators and reviewers
(\cref{sec:doi:comms:ae}),
artifact links provide a channel between creators and users
(\cref{sec:doi:comms:links}),
and artifact badges act as a channel between reviewers and users
(\cref{sec:doi:comms:badges}).
Below, we analyze the challenges associated with each of these communication channels.

\subsubsection{Artifact Evaluation and Paper Review}
\label{sec:doi:comms:ae}

Artifact evaluation provides a communication channel between creators and
reviewers that serves to assess the quality, claims, and usability of artifacts.
The effectiveness of artifact evaluation is hindered, in part, due to the lack
of open communication between creators and reviewers
\challengeref{C8: Double-blind review; C12: Limited communication}.
Despite the best efforts of authors to anticipate potential issues,
tacit knowledge and unidentified assumptions can complicate or prevent
the review process \challengeref{C2: Portability; C4: Tacit knowledge}.
On the other end, reviewers spend considerable efforts identifying and addressing
these issues during evaluation, often with little or no assistance from creators,
leaving them with little time for more than a superficial evaluation.

The absence of clear guidelines and expectations around artifacts and
artifact review represent a shared source of confusion and frustration between creators and reviewers
\challengeref{C6: Lack of standards and guidelines}.
P128 shares that their
\enquote{artifact [was] rejected because of wrong expectations about the nature
of the artifact,}
and P61 says,
\enquote{[s]ome good artifacts are rejected due to unclear standard
  of artifacts.}
Such negative experiences can deter authors from
participating in artifact evaluation entirely, thereby
undermining the reason for the existence of \acp{aec}.
For example, P31, says that
\enquote{based on wildly different reviewer expectations between
  different artifact evaluation committees, I have decided to not waste time to
  prepare a formal artifact package recently, but have focused instead on open
  source releases.}

In cases where there is no formal \ac{aec}, authors may optionally include artifacts
during paper submission.
However, authors identify that sharing artifacts as part of a double-blind
paper submission, rather than artifact evaluation, introduces the additional challenges
of removing identifying information and finding a suitable place to anonymously
host the artifact \challengeref{C7: Hosting; C8: Double-blind review}.
In the case that authors opt to provide an artifact with their submission, P64
points out that it is \enquote{not obvious to me if even the reviewers look at these artifacts when
evaluating the paper.}

\subsubsection{Artifact Links}
\label{sec:doi:comms:links}

Artifacts are typically distributed by their creators to potential users by a
URL that is included in the associated publication
\challengeref{C7: Hosting}.
Users point out that artifacts
are often provided with little or no documentation about
their contents,
how to install and use them,
and how they support the associated submission
\challengeref{C4: Tacit knowledge; C5: Artifact does not fit purpose}.
These difficulties, stemming from a lack of documentation and mismatched expectations,
lead potential users to abandon their efforts to use artifacts. 

Creators are unsure of how their artifacts should be packaged and what
documentation should be provided to avoid these difficulties.
This is partly due to inherent challenge of tacit knowledge, but is exacerbated
by the lack of broadly accepted community-wide standards, guidelines, and norms around what artifacts
should have, be, and do
\challengeref{C4: Tacit knowledge; C6: Lack of standards and guidelines}.

\subsubsection{Artifact Badges}
\label{sec:doi:comms:badges}

Artifact badges serve as a means of signaling the existence, quality, and reusability
of an artifact to prospective users.
According to \cite{Shriram15}, the presence of \acp{aec}
\enquote{sends a message that artifacts are valued and are an important part of the contribution of papers,}
\enquote{encourages authors to produce reusable
artifacts, which are the cornerstone of
future research,}
and takes the research community closer to
\enquote{the point where any published idea
that has been evaluated, measured, or
benchmarked is accompanied by the
artifact that embodies it.}

However, our survey participants complain that the standards of artifact review
are often unclear and vary considerably between different conferences and even
editions of the same conference \challengeref{C6: Lack of standards and guidelines}.
This leads to ambiguity in the perceived meaning and value of artifact badges.
As a result of this ambiguity, the reliablity of artifact badges as an indicator of quality to potential users
is diminished \challengeref{C5: Artifact does not fit purpose}.

\subsection{Time}
\label{sec:interpretation:time}

Over time, artifacts may become
inaccessible by others as short-term hosting strategies fail, such
as when an artifacts hosted on a student website becomes unavailable after the
student moves on \challengeref{C3: Maintenance}.
Even in cases where artifacts are still available, those artifacts are likely to
\enquote{bitrot} over time and become more unusable as the environments in which
they are used become increasingly dissimilar to those in which they were produced
\challengeref{C2: Portability}.

Continual maintenance is needed to fight against the inevitable process of
bitrot.
However, these activities are perceived as requiring
considerable time and resources while yielding potentially diminishing returns
to both author and user \challengeref{C1: Not worth it}.
The labor and resources costs required to maintain
an artifact dissuade some authors from sharing their artifacts in the first
place.

Alternatively, authors may mitigate the effects of bitrot by using a VM,
container, or an another form of virtual environment to package their
artifacts in a predictable, ready-to-use state.
This requires upfront effort from the authors, as retroactively packaging
artifacts as VMs or containers can be challenging and may not accurately
represent the environment that was used to conduct experiments
\challengeref{C2: Portability}.

\subsection{Social System}
\label{sec:interpretation:social}

We find that the community, as a whole, benefits when high-quality
artifacts are shared, as they provide greater knowledge,
encourage transparency, and catalyze further research.
However, the traditional academic system often does not reward
artifact authors for the creation or maintenance of high-quality artifacts, which results in insufficient time and resources
devoted to artifact sharing \challengeref{C1: Not worth it}.
The lack of incentives for both creators and reviewers can often
lead to the creation of lower-quality artifacts that are not maintained
over time \challengeref{C3: Maintenance}, and a less-effective artifact evaluation process that does not
identify and address potential issues \challengeref{C10: Lack of reviewer incentives}.
This situation prevents the community from enjoying all of the full,
long-term benefits of artifacts.

Just as the creation and sharing of artifacts is perceived to be
a low-reward activity, reviewers also note a lack of
reward and recognition for their efforts, and point out that artifact review
is less valued than paper reviews. 
Since artifact review is often less valued than technical track reviewing,
artifact reviewers tend to be less experienced and have fewer resources.
This can lead to a situation in which neither the authors or reviewers
are fully invested in the process of artifact evaluation,
and when combined with the lack of communication between the two parties
\challengeref{C12: Limited communication},
may ultimately produce inconsistent reviews and
a failure to improve and ensure the quality of artifacts.
Ultimately, the lack of incentives for both authors and reviewers
lead to problems downstream for artifact users that are not
identified and addressing during review, such as
a lack of portability, incomplete documentation, and mismatched
expectations \challengeref{C2: Portability; C4: Tacit knowledge; C6: Artifact does not fit purpose}.

\section{Recommendations}
\label{sec:recommendations}

In this section, we follow IS principles to make recommendations to improve
the experience and outcomes of artifact creation, sharing, use, and review.
Our recommendations, presented in \Cref{tab:recommendations},
are derived from the challenges identified in our
primary analysis (\cref{sec:results}), based on an understanding of how
those challenges are related within the wider context of
the research community (\cref{sec:secondary}), and are targeted at
specific subpopulations.
We discuss our recommendations below.

Note that our recommendations are based on popular
implementations of \acp{aec}. If the fundamental role and purpose of \acp{aec}
changes in the future, then some recommendations may no longer apply.
For example, if artifacts are explicitly reviewed as part
of the paper review process, as we believe they should be.

\begin{landscape}
\begin{table}[t]

\renewcommand\arraystretch{1.5}
\setlength{\aboverulesep}{0pt}
\setlength{\belowrulesep}{0pt}
\rowcolors{2}{white}{lightgray!30}

\newcommand*\rot{\rotatebox{90}}
\centering
\begin{tabular}{p{1mm}p{4mm}p{90mm}p{32mm}p{50mm}}
\toprule
& & \textbf{Recommendation} & \textbf{Derived from Results} & \textbf{Supported By Literature} \\
\midrule
\gcreator & \textbf{R1} & Describe the contents, structure, and purpose of artifacts
\begin{itemize}[noitemsep,topsep=2pt,parsep=0pt,label=\textbullet,after=\vspace{-5pt}]
\item provide citation guidelines
\item consider a permissive license that ensures creators are credited
\end{itemize}%
& C1, C4, C5, C13, Communication, Social System, Innovation
&\begin{shortlist}
\item \cite{Basili07}
\item \cite{StoddenCSE09}
\item \cite{Collberg16}
\item \cite{Flittner17}
\item \cite{Rougier2017}
\item \cite{Heumuller20}
\end{shortlist}
\\
  \gcreator & \textbf{R2} & Create a self-contained artifact
\begin{itemize}[noitemsep,topsep=2pt,parsep=0pt,label=\textbullet,after=\vspace{-5pt}]
\item consider providing a VM or container image for the artifact
\item aim to produce an artifact that is amenable to review
\end{itemize}
& C2, C11, Communication &
\begin{shortlist}
\item \cite{Heumuller20}
\end{shortlist}
\\
\gcreator\newline\gadvisor & \textbf{R3} &
Establish a long-term plan for creating and sharing artifacts
\begin{itemize}[noitemsep,topsep=2pt,parsep=0pt,label=\textbullet,after=\vspace{-5pt}]
\item use a long-term archival service to store artifacts indefinitely
\item establish and state maintenance expectations
\item foster a work environment that values artifacts
\end{itemize}
& C1, C3, C7, Time, Communication
&\begin{shortlist}
\item \cite{Basili07}
\item \cite{Collberg16}
\item \cite{Smith16}
\item \cite{Rougier2017}
\item \cite{Heumuller20}
\end{shortlist}%
\\

 \greviewer & \textbf{R4} & Obtain and use a clear rubric to evaluate artifacts
\begin{itemize}[noitemsep,topsep=2pt,parsep=0pt,label=\textbullet,after=\vspace{-5pt}]
\item determine who is responsible for reviewing artifacts
\item obtain a rubric from the \ac{aec} or conference chairs
\item or else devise a rubric and include it in the review
\end{itemize}
& C6, Communication & \\
\gprocessorg & \textbf{R5} &
Align author and reviewer expectations
\begin{itemize}[noitemsep,topsep=2pt,parsep=0pt,label=\textbullet,after=\vspace{-5pt}]
\item establish and communicate the purpose of artifact evaluation
\item communicate evaluation criteria to both reviewers and authors
\item improve review consistency by building institutional knowledge
\end{itemize}
& C1, C6, Communication, Social System &
\begin{shortlist}
\item \cite{Hermann20}
\end{shortlist}%
\\
\gprocessorg & \textbf{R6} & Reduce the opportunity cost of reviewing artifacts
\begin{itemize}[noitemsep,topsep=2pt,parsep=0pt,label=\textbullet,after=\vspace{-5pt}]
\item improve communication between authors and reviewers
\item consider introducing a \enquote{Best Artifact Reviewer} award
\item consider providing compute resources to reviewers
\end{itemize}
  & C10, C11, C12, Communication, Social System &
\begin{shortlist}
\item \cite{Hermann20}
\end{shortlist}%
\\

 \gcommunity & \textbf{R7} & Recognize and fund artifact-related activities
\begin{itemize}[noitemsep,topsep=2pt,parsep=0pt,label=\textbullet,after=\vspace{-5pt}]
\item consider artifacts during hiring, reappointment, and promotion
\item consider artifacts when reviewing grant applications
\item consider providing funding to support artifact hosting, maintenance, and review
\end{itemize}
& C1, C7, Social System, Communication, Innovation, Time &
\begin{shortlist}
\item \cite{nsf-sharing-policy}
\item \cite{EPSRCpolicy}
\item \cite{H2020FAIR,H2020pilot}
\end{shortlist}%
\\
\gcommunity & \textbf{R8} & Devise a long-term strategy for artifact sharing and evaluation
\begin{itemize}[noitemsep,topsep=2pt,parsep=0pt,label=\textbullet,after=\vspace{-5pt}]
\item establish a common artifact description format
\end{itemize}
& C1, C4, C6, C10, Social System, Communication &
\begin{shortlist}
\item \cite{Hermann20}
\end{shortlist}%
\\
\bottomrule
\end{tabular}
\caption{An overview of our recommendations,
  annotated with the challenges that they tackle and their relevant \ac{doi} dimensions,
  and a list of supporting studies and aligned proposals from the literature.
  Recommendations are targeted at creators (\gcreator), advisors and mentors (\gadvisor),
  reviewers (\greviewer), process organizers (\gprocessorg), and community leaders
  (\gcommunity).}
\label{tab:recommendations}
\end{table}
\end{landscape}

\subsection{R1: Describe the contents, structure, and purpose of artifacts}
\label{sec:recommendations:r1}

Many of the issues experienced by artifact users and reviewers ultimately stem
from mismatched expectations over what the artifact should have, be, and do.
To minimize such issues, creators should improve communication by clearly
documenting (e.g., via a README) the important details of their artifacts.
Below, we take inspiration from the ideas of
sharing contracts~\citep{Collberg16},
meta-artifacts~\citep{Flittner17},
and data sharing agreements~\citep{Basili07},
discussed in further detail in \Cref{sec:related-work},
to identify some of those important details.

For all artifacts, creators should provide a clear description of the contents,
structure and purpose of the artifact, and indicate how the artifact supports the
claims made in the associated paper.
To ensure that creators are credited for their efforts,
artifacts should be packaged with citation guidelines
that make it clear how those using and extending the artifact should cite the work.
Additionally, creators should consider using a licensing framework that encourages others to
build upon their work while ensuring that creators are credited,
such as the
Reproducible Research Standard~\citep{StoddenIJCLP09,StoddenCSE09},
Creative Commons Attribution License~\citep{cc-attribution},
and Apache License, Version 2~\citep{apachev2}.

Tool artifacts should follow existing software engineering best practices by
including details about
the intended uses, limitations, and overall generality of the tool,
and providing example uses.
Data artifacts should state the contents and format of the dataset
(e.g., columns and units of measurement), and describe
the methodology that was used to collect the data (i.e., provenance).

\subsection{R2: Create a self-contained artifact}

Creators of tool artifacts can take steps to anticipate, identify, and address
portability issues that may be encountered by users and reviewers.
At a minimum, creators should provide installation instructions,
a list of dependencies, and a description of the relevant details of
the systems on which the tool was developed and used (e.g., operating
system, machine specifications, compiler versions).
These instructions should either be checked manually, or, better yet,
automatically over the course of the project
using a continuous integration service
such as TravisCI\footnote{\url{https://travis-ci.org} [Date Accessed: March \nth{13} 2021]}
or GitHub Actions.\footnote{\url{https://github.com/features/actions} [Date Accessed: March \nth{13} 2021]}
Creators can simplify their installation process by using
popular build and package management systems
(e.g., CMake~\sqcitep{cmake},
Gradle~\sqcitep{gradle},
pip~\sqcitep{pip},
make~\sqcitep{make}),
and avoiding custom or exotic solutions.
Issues due to missing dependencies, mismatched versions,
and platform and environment incompatibilities
can be mitigated by using virtualization
(e.g., VirtualBox~\sqcitep{virtualbox},
QEMU~\sqcitep{qemu})
or containerization
(e.g., Docker,
Podman~\sqcitep{podman})
to package artifacts and run experiments
with greater reproducibility.

To maximize the effectiveness of the artifact review process,
creators should strive to produce
artifacts that are amenable for evaluation by reviewers.
This can be partly achieved by establishing expectations, stating claims,
and taking steps to avoid potential portability issues.
In the case where an artifact involves reproducing lengthy and expensive
experiments (e.g., program repair), creators should try to provide a
representative proof-of-concept form of the artifact that provides partial
evidence of its claims on a smaller dataset.

\subsection{R3: Establish a plan for creating and sharing artifacts}

At the beginning of a research project, all members of the research team,
including primary researchers, advisors, and mentors,
should establish a plan for the creation and dissemination of any
associated artifacts (e.g., tools, datasets, source code).

Plans should ensure that enough time is allocated to
artifact-related efforts (e.g., writing documentation, tests, installation instructions)
throughout the entire research project, rather than deferring activities until after
paper submission and acceptance.
As part of that effort,
advisors and mentors should work to establish a culture of valuing
artifacts within their research groups that recognizes the effort required to
produce and maintain high-quality artifacts.
Just as students are often taught how to do good research and write valuable
reviews, mentors and advisors should provide instruction and guidance as to how
to prepare and evaluate artifacts.

Plans should also be made for the long-term maintenance of the artifact.
These plans should establish expectations around how much support will be
given for an artifact, for how long, and by whom the artifact will be maintained.
Advisors, in particular, should consider making transition plans
for the case that key maintainers move onto other organizations.
To avoid mismatched expectations between creators and users, maintenance plans
should be included as part of the artifact description, together with
contact information for the artifact maintainers.

To avoid long-term availability issues, creators should
plan for the hosting and long-term archival of their artifacts.
Where possible, creators should use archival services such as
Zenodo~\citep{zenodo},
Figshare,\footnote{\url{https://figshare.com} [Date Accessed: March \nth{13} 2021]}
and Software Heritage~\citep{software-heritage}
to guarantee
the long-term availability of a snapshot of their artifacts
via a unique identifier.
Traditional file sharing services (e.g., Dropbox, Google Drive)
and temporary hosting solutions (e.g., a university website) should be
avoided as they are not designed for reliable, long-term storage.
Naturally, some artifacts (e.g., tools) evolve over time and may be used
in multiple paper submissions. VCS-based hosting services
(e.g., GitHub, GitLab,\footnote{\url{https://www.gitlab.com} [Date Accessed: March \nth{13} 2021]} BitBucket)\footnote{\url{https://www.bitbucket.org} [Date Accessed: March \nth{13} 2021]}
may be used to host the working
version of the artifact and to show its development history, but the authors should also archive a snapshot which corresponds to a specific paper using archival services.

\subsection{R4: Obtain and use a clear rubric to evaluate artifacts}

Reviewers should first establish who, if anyone, is responsible
for reviewing the artifacts associated with a paper. For conferences that have an \ac{aec},
technical reviewers should explicitly define the aspects of the artifact that
they are reviewing. For (aspects of) artifacts that they do not review, technical
reviewers may encourage authors to submit their artifacts for evaluation.
For conferences that do not have an \ac{aec}, technical reviewers may consider
reviewing artifacts, to a limited extent, themselves.

Before embarking on the review process, reviewers should ensure that they have
a clear rubric or set of criteria for evaluating artifacts.
Ideally, this should come from the conference chairs, \ac{aec} chairs,
or journal editors, and be available to authors ahead of paper submission,
reviewers prior to reviewing, and all consumers of accepted papers.
Reviewers should become familiar with the rubric before reviewing,
and apply it consistently. If they do not have access to a set of criteria,
they should ask the \ac{aec} chair, conference chair, or journal editor
if there is one available, or collaborate with other committee members to
create one.
In cases where that is not possible, reviewers should devise their own rubric,
apply it consistently during review, and communicate the criteria that was
used to evaluate the artifact to its creators.
To enhance the overall effectiveness of the review process,
reviewers may also consider using their position to ask the conference, \ac{aec}, or journal
to develop criteria for evaluating artifacts before accepting an invitation
to review.

\subsection{R5: Align author and reviewer expectations}

To align expectations between authors and
reviewers, conferences and journals should release clear and consistent
evaluation guidelines for artifacts as part of their call for papers.
These guidelines should include a description of the high-level goals and
purpose of artifact review, the extent to which artifacts
will be reviewed, and the criteria according to which artifacts will
be evaluated.
For example, guidelines should state whether reviewers should attempt to
replicate the same result as the original experiment, or if they should attempt to explore the effects of
changing parameters and/or using different data.

Authors point out that there is not enough time to prepare artifacts.
For example, P101 shares that \enquote{time constraints made it impossible to put together a high-quality, reproducible artifact before the submission deadline.}
To encourage authors to proactively develop their artifacts and allow them
to better manage their time and efforts,
conferences should consider including artifact submission dates as part of their call
for research papers.

\cite{Hermann20} find that most \acp{aec} are composed of reviewers who
have never served on an \ac{aec} before.
The resulting lack of institutional knowledge,
coupled with a lack of guidelines and clear purpose for artifact evaluation,
can lead to inconsistent reviewing and, in some cases, discourage authors
from submitting their artifacts for evaluation.
To build institutional
knowledge and improve the quality and consistency of reviews over time,
\acp{aec} should consider having a way for reviewers to hand off information
from one \enquote{generation} to the next.
As part of this process, \acp{aec} may hold a retrospective at the end of reviewing,
and make suggestions for improving their guidelines and implementation.
Journals and conferences without a separate \ac{aec} may also consider conducting
such a retrospective and periodic evaluation of guidelines.
Consistency between artifact and paper reviews may also be
improved by allowing technical reviewers to leave confidential
remarks to the \ac{aec}.

\subsection{R6: Reduce the opportunity cost of reviewing}

Artifact evaluation is often a labor-intensive activity that, compared to
traditional technical reviewing, has fewer perceived rewards
(i.e., reviewing has a high opportunity cost).
This situation can lead to lower-quality reviews, and consequently, a
less-effective review process.
To address this situation, \acp{aec}
should take steps to reduce the burden on reviewers and create incentives.
Providing clear reviewing guidelines, building institutional knowledge,
and offering adequate compute resources (e.g., cloud compute credits)
can make it easier for reviewers to review artifacts effectively.
\acp{aec} may also consider introducing a \enquote{lazy second evaluation} where
artifacts are only assigned a second reviewer if the first has concerns about
building, running, or evaluating the artifact. Improving communication
between authors and reviewers may reduce the number of cases where a second reviewer
is needed.

The overall efficiency and effectiveness of the review process
can be improved by eliminating communication issues between authors and
reviewers. One way to both improve communication and align expectations
between authors and reviewers is to require authors to provide a separate
document that describes their artifact. This document should contain
the details outlined in \Cref{sec:recommendations:r1},
including a description of the contents, claims, and limitations of the
artifact, and how it relates to the results of the associated paper.
The artifact review process can then be used as an opportunity to \enquote{certify}
the associated document for the artifact. Users of the artifact can then
use certified documents to help them quickly establish expectations around
what a particular artifact should have, be, and do.

For some \acp{aec}, communication between authors and reviewers is one-way.
Given the inevitablity of technical difficulties, this has a potential to
create an adversarial situation that fails to thoroughly identify and address
issues that may be experienced by eventual users of the artifact.
\acp{aec} should consider using continuous, two-way communication (i.e., shepherding)
that allows reviewers
to more easily evaluate the claims of the artifact, and gives room for artifacts
to be improved. One way to implement such communication without revealing
the identity of the reviewer is through the use of
an anonymizing mailbox that allows creators to respond to reviewer questions.

Just as several conferences have recently introduced \enquote{best
reviewer} rewards, \acp{aec} may consider a similar award for artifact
reviewers. This would increase the potential reward and prestige of
artifact reviewing, and with reductions to the cost of reviewing, would
decrease the opportunity cost of reviewing.

\subsection{R7: Recognize and fund artifact-related activites}

In the words of P64,
\enquote{artifacts are useful to the community, but for purposes of academic promotion it is not clear they have any value there. Unless artifacts are considered as part of career growth I do not think it will see wide adoption.}

To better align incentives, hiring and reappointment and
promotion committees should reward those who share their artifacts consistently
with the community. To that end, organizations such as the
\ac{cra}
should consider developing best practices for incentivizing and
evaluating the impact of artifact release and sustainment for tenure and
promotion policies~(c.f., \citealt{cra-best-practices}).

Funding agencies could ask (co-)principal investigators to provide
artifact creation and sharing plans in their grant applications, and may consider
the applicants track record of sharing formally evaluated artifacts during review.
Several funding agencies, including the \ac{nsf},
\ac{nih},
\ac{epsrc},
and the \ac{ec} have already implemented explicit data management
policies to encourage the sharing and reuse of artifacts~\citep{nsf-sharing-policy,nih-sharing-policy,EPSRCpolicy}.
The NSF, for example, requires that proposals provide a two-page
data management plan (DMP) that describes
\enquote{the data, metadata, scripts used to generate the data or metadata,
experimental results, samples, physical collections, software, curriculum
materials, or other materials to be produced in the course of the project.}
These efforts could be supported by grants supplements for the
long-term maintenance and archival of artifacts.
For example,
DMPs are optional for European Union Horizon 2020 proposals, but may be used to cover
certain costs related to open access~\citep{H2020pilot,H2020FAIR}.

To tackle some of the technical obstacles of artifact reviewing,
organizations such as the ACM and IEEE should consider providing
funding for compute resources (e.g., in the form of cloud-compute credits) to conferences and journals.

\subsection{R8: Establish a long-term strategy for artifact sharing and evaluation}
Editors and steering committees have experience and influence that span multiple years,
and are responsible for overseeing the
long-term continuity and direction of their respective venue. We recommend that
these entities devise a long-term strategy for the sharing and evaluation of artifacts,
and, in the case of conferences, leave the implementation of that strategy to the \ac{aec}
chairs.

Professional organizations also play a major role in the long-term future of artifact
sharing and evaluation.
The \ac{acm}, in particular, has pioneered initiatives to enhance the visibility,
usability, and availability of artifacts, such as its
badging policy and long-term archival of artifacts via the
\ac{acm} Digital Library~\citep{ACMartifacts,ACMbadging}.
Going forward, professional organizations
should work with community leaders to develop a
standard artifact description format that communicates the important details of artifacts.

To ensure that guidance maintains relevant and reflective of the community over time,
journal editors, steering committees, and professional organizations
should periodically conduct a systematic review of artifact evaluation
to identify new challenges and disseminate best practices within
the community.

\section{Related Work}
\label{sec:related-work}

\subsection{Replicability}

The concept of software artifacts today traces its lineage back to the idea of
\enquote{laboratory packages}
which describe an
\enquote{experiment in specific terms and provides materials for
replication}~\citep{Shull02,Shull08,Brooks08,basili1999building}.
Artifacts, as we understand them, are both the materials for
reuse, repurposing, and replication, and also, in some cases, the laboratory
package itself.
There are opposing views on the importance of laboratory packages
within in the literature.
\cite{Lindvall05} find that it is cheaper for researchers to
reproduce an experiment, even with
small modifications, by using a laboratory package, rather than designing
the entire replication from scratch.
\cite{Shull08} argue that laboratory packages allow others
to inexpensively (1) ensure that a given result is reproducible and thereby increase confidence
in that result, and (2) understand the sources of variability that
affect a given result so as to understand its scope and limitations.
Shull et al. consider conceptual replications (i.e., a different
lab provides an independent implementation and conducts their own experimental
setup to confirm results) to be too expensive to be considered the norm.
This view appears to be shared by many artifact evaluation committees:
Reflecting on their experiences of running \acp{aec} at multiple conferences
over several years, \cite{Shriram15} state that \enquote{repeatability is an
inexpensive and easy test of a paper’s artifacts, and clarifies the
scientific contribution of the paper,} and that reproducibility (i.e.,
conceptual replications) \enquote{is an expensive undertaking
and not something we are advocating.}
Opposed to this view, \cite{Kitchenham08} cautions that laboratory packages allow
flaws in the original experiment design to be repeated and can be
expensive in the long run, and that exact replications are of limited use
as they cannot be used in meta-analyses.
We observe a similar spectrum of views on the role of artifacts within the
responses of our survey participants.
To that end, we encourage conferences, journals, and \acp{aec},
where applicable, to explicitly state the high-level goals and purpose of their artifact evaluation
process.

\subsection{Existing Recommendations}

A number of proposals have been made to improve artifacts:
\cite{Collberg16} propose that authors
include \enquote{sharing contracts} with their papers that give basic
information about the artifacts for that paper, including the length of time
and extent (e.g., bug fixes, feature requests) to which those
artifacts will be supported.
In the field of Software Defined Networks, \cite{Flittner17}
propose that artifacts should be accompanied by
\emph{meta-artifacts}, which describe the tools and parameters that were
used during an evaluation.
\cite{StoddenCSE09,StoddenIJCLP09} proposes the Reproducible Research Standard,
a licensing framework that promotes the sharing of artifacts and ensures
that authors are credited for their work.
Numerous public platforms have been proposed that use virtualization and
containerization to facilitate the long-term persistence and
replication of data and source code artifacts~\citep{brammer2011paper,austin2011carmen,jimenez2017popper,meng2016conducting,bugzoo,fursin2016collective}.
Coding conventions and best practices have also been proposed to make it easier
for authors to produce high-quality artifacts that can be more easily understood,
reused, and replicated by others~\citep{li2012literate,aec-guidelines,stodden2014implementing}.
The majority of these various proposals are technical in nature and are largely aimed
towards artifact creators.
We incorporate elements of these proposals into our own recommendations,
but also include sociotechnical suggestions, and make tailored recommendations
to particular subpopulations of the research community.

\cite{Carver10} proposes a set of guidelines for
reporting replications of software engineering studies,
which include describing the original study, providing the motivation
and important details of the replication, and reporting both consistent
and inconsistent results.
\cite{Carver14} perform a user study to determine the
effectiveness of the proposed guidelines and find that, overall, both
reviewers and authors view them positively, provided that they do not
prescribe specific paper outlines or exact content.
Our recommendations are aimed at the various actors responsible for
directly and indirectly creating and sharing artifacts.
We do not make recommendations to artifact users or those
performing replication studies.

\cite{Basili07} identify six important properties related to
the sharing of artifacts based on their personal experiences across several
projects: Permission, credit, feedback,
protection, collaboration, and maintenance.
Our results corroborate those data sharing properties
and identify additional challenges related to the creation,
sharing, use, and review of artifacts.
Based on those data sharing properties,
they outline the notion of data sharing agreements,
which help to establish expectations around artifacts (e.g., attribution,
maintenance, costs),
and encourage the community find a means to collect and publish such documents.
We believe that Artifact Evaluation Committees, which were not introduced
until several years after Basili et al.'s recommendations,
may be a successful vehicle
for implementing such proposals on a larger scale within the community,
and we include the ideas of data sharing agreements
in our recommendations.

\subsection{Artifact Studies}

\cite{Childers17,Childers18} investigated the incentives of submitting artifacts for
evaluation. They looked at all publications at three conferences
(ECOOP, OOPSLA, and PLDI) between 2013 and 2016 and compared the average
citation count of papers that were accepted by an \ac{aec} (AE papers)
against papers that were not (non-AE papers).
Their results, while not conclusive,
suggest that AE papers may be correlated with a slightly higher citation count.

Similarly, in a study of papers published at \ac{icse} between 2007 and 2017,
\cite{Heumuller20}
find that linking artifacts to a paper leads to a small, but statistically
significant increase in citation count.
The authors determined that approximately 76.6\% of papers describe an artifact,
but that only 48\% were linked through a link in the paper, and only 56.4\% of
those artifacts were available at the links provided.
In this study, we conducted a similar analysis of artifact availability across several venues,
including, over a more recent, five-year period (2014--2018), and observed a linear increase
in the linking and availability of artifacts over time.

\cite{Kotti20} conducted a study of data papers published at \ac{msr} between
2005 and 2018 to determine how
often, by whom, and for what purpose researchers reuse their associated artifacts.
They found that 65\% of data papers have been used in other studies, but that
those papers are cited less often than technical papers at the same conference.

\cite{Hermann20} conducted a survey of individuals who had served on an \ac{aec} between
2011 and 2019 at one of several venues to understand community expectations and perceptions
about artifact use and evaluation.
They observe a similar set of perceptions among artifact reviewers and users to those
reported in our study, and make recommendations that align with our own.

Collberg et al. analyze
601 papers published at several Computer Systems conferences and journals to
determine whether their
accompanying code artifacts, if any,
can be obtained and built by others with reasonable efforts~\citep{Collberg15,Collberg16}.
This effort generated quite a controversy among the community, and lead to an effort to examine ``Reproducibility in Computer Science'' ~\citep{examining-Reproducibility}, which found different results from the original paper's results.
We did not attempt to build or use artifacts as part of our study methodology,
but found similar challenges to those reported by Collberg et al. through an
analysis of author survey responses (e.g., dead links, missing documentation, and lacking
portability).

\cite{Shull02} describe how \enquote{tacit knowledge}
(i.e., \enquote{the transfer of experimental know-how}) can be challenging
even when good artifacts are provided.
Our survey participants report similiar challenges in using and reviewing
artifacts. We see artifact evaluation as a potential means of identifying
and addressing the presence of tacit knowledge and unstated assumptions.

\subsection{Experience Reports}

\cite{Shriram15} report their experiences of running artifact evaluation
committes for five major programming languages and software engineering
conferences between 2011 and 2014.
They highlight how, between those years, participation greatly increased,
and at OOPSLA'14, 21 of 50 accepted papers (42\%) were submitted for evaluation.
From our private correspondence with \ac{aec} chairs at \ac{fse} between 2015 and 2018,
we found that participation was generally much lower (18--33\%).
In the context of our results,
this difference suggests that the
implementation of \acp{aec} within software engineering has room for improvement.
To that end, we outline recommendations to enhance the
effectiveness of \acp{aec}.

\section{Conclusion}

In this paper, we conducted a mixed-methods study to understand how
researchers create, share, and use artifacts that accompany
research papers (e.g., tools, source code, data), and the challenges that
prevent the community from realizing the full benefits of those artifacts.

We find that artifact sharing is an established norm within the
software engineering research community, and that an increasing majority of
research papers published at \ac{ase},
\ac{fse}, \ac{icse}, and \ac{emse} between 2014 and 2018 included an artifact.
Artifacts are highly valuable to the community as a whole,
but the act of creating and sharing artifacts is perceived to be a
poor investment of time that yields relatively few career benefits
compared to paper writing.
The lack of incentives for creators, coupled with the
technical challenges of creating and maintaining artifacts (e.g., hosting and portability)
and a lack of community standards and expectations around artifacts,
hinder the production of high-quality artifacts.
As a result, potential users must overcome related challenges to use artifacts,
and often need perform modifications to the artifact to fit their needs and
expectations.

\acp{aec} are a promising mechanism for preemptively identifying and addressing artifact
usability concerns.
However, we observe that
relatively few papers that contain artifacts are submitted for evaluation.
In cases where artifacts are submitted, the efficiency and effectiveness of
the evaluation process is hampered by several confounding challenges, including limited
communication between creators and reviewers, missing documentation,
and a lack of institutional knowledge.

We propose several recommendations, derived from our results and existing
proposals from the literature, to raise the quality of artifacts
and enhance the effectiveness and efficiency of artifact review.
Following IS principles, we tailor our recommendations to specific groups
based on a understanding of how artifact creation, sharing, use, and review
takes place within the community over time.

In future work, we plan to work with process organizers (e.g., \ac{aec} chairs)
to develop, evaluate, and revise evidence-based interventions to improve
the effectiveness of artifact review using \ac{is} principles.
We also plan to work with community leaders to disseminate identified best
practices to all members of the research community.

\bibliographystyle{ACM-Reference-Format}
\bibliography{bibliography.bib}


\begin{thebibliography}{81}


\ifx \showCODEN    \undefined \def \showCODEN     #1{\unskip}     \fi
\ifx \showDOI      \undefined \def \showDOI       #1{#1}\fi
\ifx \showISBNx    \undefined \def \showISBNx     #1{\unskip}     \fi
\ifx \showISBNxiii \undefined \def \showISBNxiii  #1{\unskip}     \fi
\ifx \showISSN     \undefined \def \showISSN      #1{\unskip}     \fi
\ifx \showLCCN     \undefined \def \showLCCN      #1{\unskip}     \fi
\ifx \shownote     \undefined \def \shownote      #1{#1}          \fi
\ifx \showarticletitle \undefined \def \showarticletitle #1{#1}   \fi
\ifx \showURL      \undefined \def \showURL       {\relax}        \fi
\providecommand\bibfield[2]{#2}
\providecommand\bibinfo[2]{#2}
\providecommand\natexlab[1]{#1}
\providecommand\showeprint[2][]{arXiv:#2}

\bibitem[\protect\citeauthoryear{{Apache}}{{Apache}}{2004}]%
        {apachev2}
\bibfield{author}{\bibinfo{person}{{Apache}}.} \bibinfo{year}{2004}\natexlab{}.
\newblock \bibinfo{title}{{Apache License, Version 2.0}}.
\newblock
\newblock
\urldef\tempurl%
\url{https://www.apache.org/licenses/LICENSE-2.0}
\showURL{%
\tempurl}
\newblock
\shownote{[Date Accessed: July 21st 2020].}


\bibitem[\protect\citeauthoryear{{Association for Computing
  Machinery}}{{Association for Computing Machinery}}{2018}]%
        {ACMbadging}
\bibfield{author}{\bibinfo{person}{{Association for Computing Machinery}}.}
  \bibinfo{year}{2018}\natexlab{}.
\newblock \bibinfo{title}{{Artifact Review and Badging}}.
\newblock
\newblock
\urldef\tempurl%
\url{https://www.acm.org/publications/policies/artifact-review-badging}
\showURL{%
\tempurl}
\newblock
\shownote{[Date Accessed: March 30th 2020].}


\bibitem[\protect\citeauthoryear{{Association for Computing
  Machinery}}{{Association for Computing Machinery}}{2020}]%
        {ACMartifacts}
\bibfield{author}{\bibinfo{person}{{Association for Computing Machinery}}.}
  \bibinfo{year}{2020}\natexlab{}.
\newblock \bibinfo{title}{{Software and Data Artifacts in the ACM Digital
  Library}}.
\newblock
\newblock
\urldef\tempurl%
\url{https://www.acm.org/publications/artifacts}
\showURL{%
\tempurl}
\newblock
\shownote{[Date Accessed: July 29th 2020].}


\bibitem[\protect\citeauthoryear{Austin, Jackson, Fletcher, Jessop, Liang,
  Weeks, Smith, Ingram, and Watson}{Austin et~al\mbox{.}}{2011}]%
        {austin2011carmen}
\bibfield{author}{\bibinfo{person}{J. Austin}, \bibinfo{person}{T. Jackson},
  \bibinfo{person}{M. Fletcher}, \bibinfo{person}{M. Jessop},
  \bibinfo{person}{B. Liang}, \bibinfo{person}{M. Weeks}, \bibinfo{person}{L.
  Smith}, \bibinfo{person}{C. Ingram}, {and} \bibinfo{person}{P. Watson}.}
  \bibinfo{year}{2011}\natexlab{}.
\newblock \showarticletitle{{CARMEN: code analysis, repository and modeling for
  e-neuroscience}}.
\newblock \bibinfo{journal}{\emph{Procedia Computer Science}}
  \bibinfo{volume}{4} (\bibinfo{year}{2011}), \bibinfo{pages}{768--777}.
\newblock


\bibitem[\protect\citeauthoryear{Basili, Shull, and Lanubile}{Basili
  et~al\mbox{.}}{1999}]%
        {basili1999building}
\bibfield{author}{\bibinfo{person}{V.~R. Basili}, \bibinfo{person}{F. Shull},
  {and} \bibinfo{person}{F. Lanubile}.} \bibinfo{year}{1999}\natexlab{}.
\newblock \showarticletitle{Building knowledge through families of
  experiments}.
\newblock \bibinfo{journal}{\emph{Transactions on Software Engineering}}
  \bibinfo{volume}{25}, \bibinfo{number}{4} (\bibinfo{year}{1999}),
  \bibinfo{pages}{456--473}.
\newblock


\bibitem[\protect\citeauthoryear{Basili, Zelkowitz, Sj{\o}berg, Johnson, and
  Cowling}{Basili et~al\mbox{.}}{2007}]%
        {Basili07}
\bibfield{author}{\bibinfo{person}{V.~R. Basili}, \bibinfo{person}{M.~V.
  Zelkowitz}, \bibinfo{person}{D.~I.~K. Sj{\o}berg}, \bibinfo{person}{P.
  Johnson}, {and} \bibinfo{person}{A.~J. Cowling}.}
  \bibinfo{year}{2007}\natexlab{}.
\newblock \showarticletitle{Protocols in the use of empirical software
  engineering artifacts}.
\newblock \bibinfo{journal}{\emph{Empirical Software Engineering}}
  \bibinfo{volume}{12}, \bibinfo{number}{1} (\bibinfo{date}{01 Feb}
  \bibinfo{year}{2007}), \bibinfo{pages}{107--119}.
\newblock
\showISSN{1573-7616}


\bibitem[\protect\citeauthoryear{Bauer, Damschroder, Hagedorn, Smith, and
  Kilbourne}{Bauer et~al\mbox{.}}{2015}]%
        {bauer2015introduction}
\bibfield{author}{\bibinfo{person}{M.~S. Bauer}, \bibinfo{person}{L.
  Damschroder}, \bibinfo{person}{H. Hagedorn}, \bibinfo{person}{J. Smith},
  {and} \bibinfo{person}{A.~M. Kilbourne}.} \bibinfo{year}{2015}\natexlab{}.
\newblock \showarticletitle{An introduction to implementation science for the
  non-specialist}.
\newblock \bibinfo{journal}{\emph{BMC Psychology}} \bibinfo{volume}{3},
  \bibinfo{number}{1} (\bibinfo{year}{2015}), \bibinfo{pages}{32}.
\newblock


\bibitem[\protect\citeauthoryear{Beller}{Beller}{2020}]%
        {beller-aec-blog}
\bibfield{author}{\bibinfo{person}{M. Beller}.}
  \bibinfo{year}{2020}\natexlab{}.
\newblock \bibinfo{title}{{Why I will never join an Artifacts Evaluation
  Committee Again}}.
\newblock
\newblock
\urldef\tempurl%
\url{https://inventitech.com/blog/why-i-will-never-review-artifacts-again}
\showURL{%
\tempurl}
\newblock
\shownote{[Date Accessed: July 16th 2020].}


\bibitem[\protect\citeauthoryear{Brammer, Crosby, Matthews, and
  Williams}{Brammer et~al\mbox{.}}{2011}]%
        {brammer2011paper}
\bibfield{author}{\bibinfo{person}{G.~R. Brammer}, \bibinfo{person}{R.~W.
  Crosby}, \bibinfo{person}{S.~J. Matthews}, {and} \bibinfo{person}{T.~L.
  Williams}.} \bibinfo{year}{2011}\natexlab{}.
\newblock \showarticletitle{Paper m{\^a}ch{\'e}: Creating dynamic reproducible
  science}.
\newblock \bibinfo{journal}{\emph{Procedia Computer Science}}
  \bibinfo{volume}{4} (\bibinfo{year}{2011}), \bibinfo{pages}{658--667}.
\newblock


\bibitem[\protect\citeauthoryear{Brooks, Roper, Wood, Daly, and Miller}{Brooks
  et~al\mbox{.}}{2008}]%
        {Brooks08}
\bibfield{author}{\bibinfo{person}{A. Brooks}, \bibinfo{person}{M. Roper},
  \bibinfo{person}{M. Wood}, \bibinfo{person}{J. Daly}, {and}
  \bibinfo{person}{J. Miller}.} \bibinfo{year}{2008}\natexlab{}.
\newblock \bibinfo{booktitle}{\emph{{Replication's Role in Software
  Engineering}}}.
\newblock \bibinfo{publisher}{Springer London}, \bibinfo{pages}{365--379}.
\newblock


\bibitem[\protect\citeauthoryear{Carver}{Carver}{2010}]%
        {Carver10}
\bibfield{author}{\bibinfo{person}{J.~C. Carver}.}
  \bibinfo{year}{2010}\natexlab{}.
\newblock \showarticletitle{Towards reporting guidelines for experimental
  replications: A proposal}. In \bibinfo{booktitle}{\emph{{International
  Workshop on Replication in Empirical Software Engineering Research}}}
  \emph{(\bibinfo{series}{RESER '10})}.
\newblock


\bibitem[\protect\citeauthoryear{Carver, Juristo, Baldassarre, and
  Vegas}{Carver et~al\mbox{.}}{2014}]%
        {Carver14}
\bibfield{author}{\bibinfo{person}{J.~C. Carver}, \bibinfo{person}{N. Juristo},
  \bibinfo{person}{M.~T. Baldassarre}, {and} \bibinfo{person}{S. Vegas}.}
  \bibinfo{year}{2014}\natexlab{}.
\newblock \showarticletitle{Replications of software engineering experiments}.
\newblock \bibinfo{journal}{\emph{Empirical Software Engineering}}
  \bibinfo{volume}{19}, \bibinfo{number}{2} (\bibinfo{date}{01 Apr}
  \bibinfo{year}{2014}), \bibinfo{pages}{267--276}.
\newblock


\bibitem[\protect\citeauthoryear{Charmaz}{Charmaz}{2014}]%
        {charmaz2014constructing}
\bibfield{author}{\bibinfo{person}{K. Charmaz}.}
  \bibinfo{year}{2014}\natexlab{}.
\newblock \bibinfo{booktitle}{\emph{Constructing grounded theory}}.
\newblock \bibinfo{publisher}{sage}.
\newblock


\bibitem[\protect\citeauthoryear{Childers and Chrysanthis}{Childers and
  Chrysanthis}{2017}]%
        {Childers17}
\bibfield{author}{\bibinfo{person}{B.~R. Childers} {and} \bibinfo{person}{P.~K.
  Chrysanthis}.} \bibinfo{year}{2017}\natexlab{}.
\newblock \showarticletitle{Artifact Evaluation: Is It a Real Incentive?}. In
  \bibinfo{booktitle}{\emph{International Conference on e-Science}}
  \emph{(\bibinfo{series}{e-Science '17})}. \bibinfo{pages}{488--489}.
\newblock


\bibitem[\protect\citeauthoryear{Childers and Chrysanthis}{Childers and
  Chrysanthis}{2018}]%
        {Childers18}
\bibfield{author}{\bibinfo{person}{B.~R. Childers} {and} \bibinfo{person}{P.~K.
  Chrysanthis}.} \bibinfo{year}{2018}\natexlab{}.
\newblock \showarticletitle{Artifact Evaluation: FAD or Real News?}. In
  \bibinfo{booktitle}{\emph{International Conference on Data Engineering}}
  \emph{(\bibinfo{series}{ICDE '18})}. \bibinfo{pages}{1664--1665}.
\newblock


\bibitem[\protect\citeauthoryear{Collberg and Proebsting}{Collberg and
  Proebsting}{2016}]%
        {Collberg16}
\bibfield{author}{\bibinfo{person}{C. Collberg} {and} \bibinfo{person}{T.~A.
  Proebsting}.} \bibinfo{year}{2016}\natexlab{}.
\newblock \showarticletitle{Repeatability in Computer Systems Research}.
\newblock \bibinfo{journal}{\emph{Commun. ACM}} \bibinfo{volume}{59},
  \bibinfo{number}{3} (\bibinfo{date}{Feb.} \bibinfo{year}{2016}),
  \bibinfo{pages}{62--69}.
\newblock


\bibitem[\protect\citeauthoryear{Collberg, Proebsting, and Warren}{Collberg
  et~al\mbox{.}}{2015}]%
        {Collberg15}
\bibfield{author}{\bibinfo{person}{C. Collberg}, \bibinfo{person}{T.~A.
  Proebsting}, {and} \bibinfo{person}{A.~M. Warren}.}
  \bibinfo{year}{2015}\natexlab{}.
\newblock \bibinfo{booktitle}{\emph{{Repeatability and Benefaction in Computer
  Systems Research: A Study and Modest Proposal}}}.
\newblock \bibinfo{type}{{T}echnical {R}eport} TR 14-04.
  \bibinfo{institution}{University of Arizona}.
\newblock
\urldef\tempurl%
\url{http://reproducibility.cs.arizona.edu/v2/RepeatabilityTR.pdf}
\showURL{%
\tempurl}


\bibitem[\protect\citeauthoryear{Cosmo and Zacchiroli}{Cosmo and
  Zacchiroli}{2017}]%
        {software-heritage}
\bibfield{author}{\bibinfo{person}{R.~Di Cosmo} {and} \bibinfo{person}{S.
  Zacchiroli}.} \bibinfo{year}{2017}\natexlab{}.
\newblock \showarticletitle{{Software Heritage: Why and How to Preserve
  Software Source Code}}. In \bibinfo{booktitle}{\emph{International Conference
  on Digital Preservation}} \emph{(\bibinfo{series}{iPRES '17})}.
\newblock


\bibitem[\protect\citeauthoryear{{Creative Commons}}{{Creative
  Commons}}{2013}]%
        {cc-attribution}
\bibfield{author}{\bibinfo{person}{{Creative Commons}}.}
  \bibinfo{year}{2013}\natexlab{}.
\newblock \bibinfo{title}{{Attribution 4.0 International (CC BY 4.0)}}.
\newblock
\newblock
\urldef\tempurl%
\url{https://creativecommons.org/licenses/by/4.0}
\showURL{%
\tempurl}
\newblock
\shownote{[Date Accessed: July 21st 2020].}


\bibitem[\protect\citeauthoryear{Creswell and Clark}{Creswell and
  Clark}{2017}]%
        {creswell2017designing}
\bibfield{author}{\bibinfo{person}{J.~W. Creswell} {and}
  \bibinfo{person}{V.~L.~P. Clark}.} \bibinfo{year}{2017}\natexlab{}.
\newblock \bibinfo{booktitle}{\emph{Designing and Conducting Mixed Methods
  Research}}.
\newblock \bibinfo{publisher}{Sage Publications}.
\newblock


\bibitem[\protect\citeauthoryear{{Docker Inc.}}{{Docker Inc.}}{2020}]%
        {docker}
\bibfield{author}{\bibinfo{person}{{Docker Inc.}}}
  \bibinfo{year}{2020}\natexlab{}.
\newblock \bibinfo{title}{{Docker}}.
\newblock
\newblock
\urldef\tempurl%
\url{https://www.docker.com}
\showURL{%
\tempurl}
\newblock
\shownote{[Date Accessed: December 11th 2020].}


\bibitem[\protect\citeauthoryear{El~Emam, Jonker, Arbuckle, and Malin}{El~Emam
  et~al\mbox{.}}{2011}]%
        {el2011systematic}
\bibfield{author}{\bibinfo{person}{K. El~Emam}, \bibinfo{person}{E. Jonker},
  \bibinfo{person}{L. Arbuckle}, {and} \bibinfo{person}{B. Malin}.}
  \bibinfo{year}{2011}\natexlab{}.
\newblock \showarticletitle{A systematic review of re-identification attacks on
  health data}.
\newblock \bibinfo{journal}{\emph{PLOS One}} \bibinfo{volume}{6},
  \bibinfo{number}{12} (\bibinfo{year}{2011}), \bibinfo{pages}{e28071}.
\newblock


\bibitem[\protect\citeauthoryear{{Engineering and Physical Sciences Research
  Council}}{{Engineering and Physical Sciences Research Council}}{2011}]%
        {EPSRCpolicy}
\bibfield{author}{\bibinfo{person}{{Engineering and Physical Sciences Research
  Council}}.} \bibinfo{year}{2011}\natexlab{}.
\newblock \bibinfo{title}{{EPSRC policy framework on research data}}.
\newblock
\newblock
\urldef\tempurl%
\url{https://epsrc.ukri.org/about/standards/researchdata}
\showURL{%
\tempurl}
\newblock
\shownote{[Date Accessed: July 31st 2020].}


\bibitem[\protect\citeauthoryear{{European Commission}}{{European
  Commission}}{2016}]%
        {H2020FAIR}
\bibfield{author}{\bibinfo{person}{{European Commission}}.}
  \bibinfo{year}{2016}\natexlab{}.
\newblock \bibinfo{title}{{H2020 Programme: Guidelines on FAIR Data Management
  in Horizon 2020 (Version 3.0)}}.
\newblock
\newblock
\urldef\tempurl%
\url{https://ec.europa.eu/research/participants/data/ref/h2020/grants_manual/hi/oa_pilot/h2020-hi-oa-data-mgt_en.pdf}
\showURL{%
\tempurl}
\newblock
\shownote{[Date Accessed: July 31st 2020].}


\bibitem[\protect\citeauthoryear{{European Commission}}{{European
  Commission}}{2020}]%
        {H2020pilot}
\bibfield{author}{\bibinfo{person}{{European Commission}}.}
  \bibinfo{year}{2020}\natexlab{}.
\newblock \bibinfo{title}{{Horizon 2020: Open Access}}.
\newblock
\newblock
\urldef\tempurl%
\url{https://ec.europa.eu/research/participants/docs/h2020-funding-guide/cross-cutting-issues/open-access-data-management/open-access_en.htm}
\showURL{%
\tempurl}
\newblock
\shownote{[Date Accessed: July 31st 2020].}


\bibitem[\protect\citeauthoryear{{European Organization For Nuclear Research}
  and {OpenAIRE}}{{European Organization For Nuclear Research} and
  {OpenAIRE}}{2013}]%
        {zenodo}
\bibfield{author}{\bibinfo{person}{{European Organization For Nuclear
  Research}} {and} \bibinfo{person}{{OpenAIRE}}.}
  \bibinfo{year}{2013}\natexlab{}.
\newblock \bibinfo{title}{Zenodo}.
\newblock
\newblock
\urldef\tempurl%
\url{https://doi.org/10.25495/7GXK-RD71}
\showDOI{\tempurl}


\bibitem[\protect\citeauthoryear{Flittner, Bauer, Rizk, Gei{\ss}ler, Zinner,
  and Zitterbart}{Flittner et~al\mbox{.}}{2017}]%
        {Flittner17}
\bibfield{author}{\bibinfo{person}{M. Flittner}, \bibinfo{person}{R. Bauer},
  \bibinfo{person}{A. Rizk}, \bibinfo{person}{S. Gei{\ss}ler},
  \bibinfo{person}{T. Zinner}, {and} \bibinfo{person}{M. Zitterbart}.}
  \bibinfo{year}{2017}\natexlab{}.
\newblock \showarticletitle{Taming the Complexity of Artifact Reproducibility}.
  In \bibinfo{booktitle}{\emph{Reproducibility Workshop}}
  \emph{(\bibinfo{series}{Reproducibility '17})}. \bibinfo{pages}{14--16}.
\newblock


\bibitem[\protect\citeauthoryear{Frambach and Schillewaert}{Frambach and
  Schillewaert}{2002}]%
        {frambach2002}
\bibfield{author}{\bibinfo{person}{R.~T. Frambach} {and} \bibinfo{person}{N.
  Schillewaert}.} \bibinfo{year}{2002}\natexlab{}.
\newblock \showarticletitle{Organizational innovation adoption: A multi-level
  framework of determinants and opportunities for future research}.
\newblock \bibinfo{journal}{\emph{Journal of Business Research}}
  \bibinfo{volume}{55}, \bibinfo{number}{2} (\bibinfo{year}{2002}),
  \bibinfo{pages}{163--176}.
\newblock


\bibitem[\protect\citeauthoryear{{Free Software Foundation}}{{Free Software
  Foundation}}{2021}]%
        {make}
\bibfield{author}{\bibinfo{person}{{Free Software Foundation}}.}
  \bibinfo{year}{2021}\natexlab{}.
\newblock \bibinfo{title}{{GNU Make}}.
\newblock
\newblock
\urldef\tempurl%
\url{https://www.gnu.org/software/make}
\showURL{%
\tempurl}
\newblock
\shownote{[Date Accessed: March \nth{13} 2021].}


\bibitem[\protect\citeauthoryear{Fursin, Lokhmotov, and Plowman}{Fursin
  et~al\mbox{.}}{2016}]%
        {fursin2016collective}
\bibfield{author}{\bibinfo{person}{G. Fursin}, \bibinfo{person}{A. Lokhmotov},
  {and} \bibinfo{person}{E. Plowman}.} \bibinfo{year}{2016}\natexlab{}.
\newblock \showarticletitle{Collective Knowledge: towards R\&D sustainability}.
  In \bibinfo{booktitle}{\emph{Design, Automation \& Test in Europe Conference
  \& Exhibition}} \emph{(\bibinfo{series}{DATE '16})}.
  \bibinfo{pages}{864--869}.
\newblock


\bibitem[\protect\citeauthoryear{Glasgow, Vogt, and Boles}{Glasgow
  et~al\mbox{.}}{1999}]%
        {glasgow1999evaluating}
\bibfield{author}{\bibinfo{person}{R.~E. Glasgow}, \bibinfo{person}{T.~M.
  Vogt}, {and} \bibinfo{person}{S.~M. Boles}.} \bibinfo{year}{1999}\natexlab{}.
\newblock \showarticletitle{Evaluating the public health impact of health
  promotion interventions: the RE-AIM framework}.
\newblock \bibinfo{journal}{\emph{American Journal of Public Health}}
  \bibinfo{volume}{89}, \bibinfo{number}{9} (\bibinfo{year}{1999}),
  \bibinfo{pages}{1322--1327}.
\newblock


\bibitem[\protect\citeauthoryear{G{\'o}mez, Cleary, and Singer}{G{\'o}mez
  et~al\mbox{.}}{2013}]%
        {gomez2013study}
\bibfield{author}{\bibinfo{person}{C. G{\'o}mez}, \bibinfo{person}{B. Cleary},
  {and} \bibinfo{person}{L. Singer}.} \bibinfo{year}{2013}\natexlab{}.
\newblock \showarticletitle{A study of innovation diffusion through link
  sharing on stack overflow}. In \bibinfo{booktitle}{\emph{Working Conference
  on Mining Software Repositories}} \emph{(\bibinfo{series}{MSR '13})}.
  \bibinfo{pages}{81--84}.
\newblock


\bibitem[\protect\citeauthoryear{{Gradle}}{{Gradle}}{2021}]%
        {gradle}
\bibfield{author}{\bibinfo{person}{{Gradle}}.} \bibinfo{year}{2021}\natexlab{}.
\newblock \bibinfo{title}{{Gradle Build Tool}}.
\newblock
\newblock
\urldef\tempurl%
\url{https://gradle.org}
\showURL{%
\tempurl}
\newblock
\shownote{[Date Accessed: March \nth{13} 2021].}


\bibitem[\protect\citeauthoryear{Green and Hevner}{Green and Hevner}{2000}]%
        {green2000successful}
\bibfield{author}{\bibinfo{person}{G.~C. Green} {and} \bibinfo{person}{A.~R.
  Hevner}.} \bibinfo{year}{2000}\natexlab{}.
\newblock \showarticletitle{The successful diffusion of innovations: guidance
  for software development organizations}.
\newblock \bibinfo{journal}{\emph{IEEE Software}} \bibinfo{volume}{17},
  \bibinfo{number}{6} (\bibinfo{year}{2000}), \bibinfo{pages}{96--103}.
\newblock


\bibitem[\protect\citeauthoryear{Hager}{Hager}{2016}]%
        {pdfx}
\bibfield{author}{\bibinfo{person}{C. Hager}.} \bibinfo{year}{2016}\natexlab{}.
\newblock \bibinfo{title}{{PDFx}}.
\newblock
\newblock
\urldef\tempurl%
\url{https://github.com/metachris/pdfx/releases/tag/v1.3.0}
\showURL{%
\tempurl}
\newblock
\shownote{[Date Accessed: March \nth{28} 2021].}


\bibitem[\protect\citeauthoryear{Herckis}{Herckis}{2018}]%
        {herckis2018passing}
\bibfield{author}{\bibinfo{person}{L. Herckis}.}
  \bibinfo{year}{2018}\natexlab{}.
\newblock \showarticletitle{Passing the baton: Digital literacy and sustained
  implementation of elearning technologies}.
\newblock \bibinfo{journal}{\emph{{Current Issues in Emerging eLearning}}}
  \bibinfo{volume}{5}, \bibinfo{number}{1} (\bibinfo{year}{2018}),
  \bibinfo{pages}{4}.
\newblock


\bibitem[\protect\citeauthoryear{Hermann, Winter, and Siegmund}{Hermann
  et~al\mbox{.}}{2020}]%
        {Hermann20}
\bibfield{author}{\bibinfo{person}{B. Hermann}, \bibinfo{person}{S. Winter},
  {and} \bibinfo{person}{J. Siegmund}.} \bibinfo{year}{2020}\natexlab{}.
\newblock \showarticletitle{Community Expectations for Research Artifacts and
  Evaluation Processes}. In \bibinfo{booktitle}{\emph{Joint Meeting on European
  Software Engineering Conference and Symposium on the Foundations of Software
  Engineering}} \emph{(\bibinfo{series}{ESEC/FSE '20})}.
  \bibinfo{pages}{469--480}.
\newblock


\bibitem[\protect\citeauthoryear{Heum{\"u}ller, Nielebock, Kr{\"u}ger, and
  Ortmeier}{Heum{\"u}ller et~al\mbox{.}}{2020}]%
        {Heumuller20}
\bibfield{author}{\bibinfo{person}{R. Heum{\"u}ller}, \bibinfo{person}{S.
  Nielebock}, \bibinfo{person}{J. Kr{\"u}ger}, {and} \bibinfo{person}{F.
  Ortmeier}.} \bibinfo{year}{2020}\natexlab{}.
\newblock \showarticletitle{Publish or Perish, but do not Forget your Software
  Artifacts}.
\newblock \bibinfo{journal}{\emph{Empirical Software Engineering}}
  \bibinfo{volume}{25}, \bibinfo{number}{6} (\bibinfo{year}{2020}),
  \bibinfo{pages}{4585--4616}.
\newblock


\bibitem[\protect\citeauthoryear{Ji, Li, Srivatsa, and Beyah}{Ji
  et~al\mbox{.}}{2014}]%
        {ji2014structural}
\bibfield{author}{\bibinfo{person}{S. Ji}, \bibinfo{person}{W. Li},
  \bibinfo{person}{M. Srivatsa}, {and} \bibinfo{person}{R. Beyah}.}
  \bibinfo{year}{2014}\natexlab{}.
\newblock \showarticletitle{Structural data de-anonymization: Quantification,
  practice, and implications}. In \bibinfo{booktitle}{\emph{Conference on
  Computer and Communications Security}} \emph{(\bibinfo{series}{SIGSAC '14})}.
  \bibinfo{pages}{1040--1053}.
\newblock


\bibitem[\protect\citeauthoryear{Jimenez, Sevilla, Watkins, Maltzahn, Lofstead,
  Mohror, Arpaci-Dusseau, and Arpaci-Dusseau}{Jimenez et~al\mbox{.}}{2017}]%
        {jimenez2017popper}
\bibfield{author}{\bibinfo{person}{I. Jimenez}, \bibinfo{person}{M. Sevilla},
  \bibinfo{person}{N. Watkins}, \bibinfo{person}{C. Maltzahn},
  \bibinfo{person}{J. Lofstead}, \bibinfo{person}{K. Mohror},
  \bibinfo{person}{A. Arpaci-Dusseau}, {and} \bibinfo{person}{R.
  Arpaci-Dusseau}.} \bibinfo{year}{2017}\natexlab{}.
\newblock \showarticletitle{The Popper convention: Making reproducible systems
  evaluation practical}. In \bibinfo{booktitle}{\emph{International Parallel
  and Distributed Processing Symposium Workshops}}
  \emph{(\bibinfo{series}{IPDPSW '17})}. \bibinfo{pages}{1561--1570}.
\newblock


\bibitem[\protect\citeauthoryear{Johns}{Johns}{1993}]%
        {johns1993constraints}
\bibfield{author}{\bibinfo{person}{G. Johns}.} \bibinfo{year}{1993}\natexlab{}.
\newblock \showarticletitle{Constraints on the adoption of psychology-based
  personnel practices: Lessons from organizational innovation}.
\newblock \bibinfo{journal}{\emph{Personnel Psychology}} \bibinfo{volume}{46},
  \bibinfo{number}{3} (\bibinfo{year}{1993}), \bibinfo{pages}{569--592}.
\newblock


\bibitem[\protect\citeauthoryear{Kitchenham}{Kitchenham}{2008}]%
        {Kitchenham08}
\bibfield{author}{\bibinfo{person}{B. Kitchenham}.}
  \bibinfo{year}{2008}\natexlab{}.
\newblock \showarticletitle{The role of replications in empirical software
  engineering—a word of warning}.
\newblock \bibinfo{journal}{\emph{Empirical Software Engineering}}
  \bibinfo{volume}{13} (\bibinfo{date}{04} \bibinfo{year}{2008}),
  \bibinfo{pages}{219--221}.
\newblock


\bibitem[\protect\citeauthoryear{Kitto, Chesters, and Grbich}{Kitto
  et~al\mbox{.}}{2008}]%
        {kitto2008quality}
\bibfield{author}{\bibinfo{person}{S.~C. Kitto}, \bibinfo{person}{J. Chesters},
  {and} \bibinfo{person}{C. Grbich}.} \bibinfo{year}{2008}\natexlab{}.
\newblock \showarticletitle{Quality in qualitative research}.
\newblock \bibinfo{journal}{\emph{Medical Journal of Australia}}
  \bibinfo{volume}{188}, \bibinfo{number}{4} (\bibinfo{year}{2008}),
  \bibinfo{pages}{243--246}.
\newblock


\bibitem[\protect\citeauthoryear{{Kitware}}{{Kitware}}{2021}]%
        {cmake}
\bibfield{author}{\bibinfo{person}{{Kitware}}.}
  \bibinfo{year}{2021}\natexlab{}.
\newblock \bibinfo{title}{{CMake}}.
\newblock
\newblock
\urldef\tempurl%
\url{https://cmake.org}
\showURL{%
\tempurl}
\newblock
\shownote{[Date Accessed: March \nth{13} 2021].}


\bibitem[\protect\citeauthoryear{Kluyver, Ragan-Kelley, P{\'e}rez, Granger,
  Bussonnier, Frederic, Kelley, Hamrick, Grout, Corlay, Ivanov, Avila, Abdalla,
  Willing, and {Jupyter Development Team}}{Kluyver et~al\mbox{.}}{2016}]%
        {jupyter}
\bibfield{author}{\bibinfo{person}{T. Kluyver}, \bibinfo{person}{B.
  Ragan-Kelley}, \bibinfo{person}{F. P{\'e}rez}, \bibinfo{person}{B. Granger},
  \bibinfo{person}{M. Bussonnier}, \bibinfo{person}{J. Frederic},
  \bibinfo{person}{K. Kelley}, \bibinfo{person}{J. Hamrick},
  \bibinfo{person}{J. Grout}, \bibinfo{person}{S. Corlay}, \bibinfo{person}{P.
  Ivanov}, \bibinfo{person}{D. Avila}, \bibinfo{person}{S. Abdalla},
  \bibinfo{person}{C. Willing}, {and} \bibinfo{person}{{Jupyter Development
  Team}}.} \bibinfo{year}{2016}\natexlab{}.
\newblock \showarticletitle{Jupyter Notebooks - a publishing format for
  reproducible computational workflows}. In
  \bibinfo{booktitle}{\emph{Positioning and Power in Academic Publishing:
  Players, Agents and Agendas}}. \bibinfo{publisher}{IOS Press},
  \bibinfo{pages}{87--90}.
\newblock


\bibitem[\protect\citeauthoryear{{Kotti}, {Kravvaritis}, {Dritsa}, and
  {Spinellis}}{{Kotti} et~al\mbox{.}}{2020}]%
        {Kotti20}
\bibfield{author}{\bibinfo{person}{Z. {Kotti}}, \bibinfo{person}{K.
  {Kravvaritis}}, \bibinfo{person}{K. {Dritsa}}, {and} \bibinfo{person}{D.
  {Spinellis}}.} \bibinfo{year}{2020}\natexlab{}.
\newblock \showarticletitle{{Standing on shoulders or feet? An extended study
  on the usage of the MSR data papers}}.
\newblock \bibinfo{journal}{\emph{{Empirical Software Engineering}}}
  \bibinfo{volume}{25}, \bibinfo{number}{5} (\bibinfo{year}{2020}),
  \bibinfo{pages}{3288--3322}.
\newblock


\bibitem[\protect\citeauthoryear{Krishnamurthi}{Krishnamurthi}{2013a}]%
        {Shriram13}
\bibfield{author}{\bibinfo{person}{S. Krishnamurthi}.}
  \bibinfo{year}{2013}\natexlab{a}.
\newblock \showarticletitle{Artifact Evaluation for Software Conferences}.
\newblock \bibinfo{journal}{\emph{SIGSOFT Software Engineering Notes}}
  \bibinfo{volume}{38}, \bibinfo{number}{3} (\bibinfo{date}{May}
  \bibinfo{year}{2013}), \bibinfo{pages}{7--10}.
\newblock
\showISSN{0163-5948}


\bibitem[\protect\citeauthoryear{Krishnamurthi}{Krishnamurthi}{2013b}]%
        {examining-Reproducibility}
\bibfield{author}{\bibinfo{person}{S. Krishnamurthi}.}
  \bibinfo{year}{2013}\natexlab{b}.
\newblock \bibinfo{title}{{Examining ``Reproducibility in Computer Science''}}.
\newblock
\newblock
\urldef\tempurl%
\url{http://cs.brown.edu/~sk/Memos/Examining-Reproducibility/}
\showURL{%
\tempurl}
\newblock
\shownote{[Date Accessed: January 6th 2020].}


\bibitem[\protect\citeauthoryear{Krishnamurthi}{Krishnamurthi}{2014}]%
        {aec-guidelines}
\bibfield{author}{\bibinfo{person}{S. Krishnamurthi}.}
  \bibinfo{year}{2014}\natexlab{}.
\newblock \bibinfo{title}{{Guidelines for Packaging AEC Submissions}}.
\newblock
\newblock
\urldef\tempurl%
\url{https://www.artifact-eval.org/guidelines.html}
\showURL{%
\tempurl}
\newblock
\shownote{[Date Accessed: July 17th 2020].}


\bibitem[\protect\citeauthoryear{Krishnamurthi and Vitek}{Krishnamurthi and
  Vitek}{2015}]%
        {Shriram15}
\bibfield{author}{\bibinfo{person}{S. Krishnamurthi} {and} \bibinfo{person}{J.
  Vitek}.} \bibinfo{year}{2015}\natexlab{}.
\newblock \showarticletitle{The Real Software Crisis: Repeatability As a Core
  Value}.
\newblock \bibinfo{journal}{\emph{Commun. ACM}} \bibinfo{volume}{58},
  \bibinfo{number}{3} (\bibinfo{date}{Feb.} \bibinfo{year}{2015}),
  \bibinfo{pages}{34--36}.
\newblock


\bibitem[\protect\citeauthoryear{Li-Thiao-T{\'e}}{Li-Thiao-T{\'e}}{2012}]%
        {li2012literate}
\bibfield{author}{\bibinfo{person}{S. Li-Thiao-T{\'e}}.}
  \bibinfo{year}{2012}\natexlab{}.
\newblock \showarticletitle{Literate program execution for reproducible
  research and executable papers}.
\newblock \bibinfo{journal}{\emph{Procedia Computer Science}}
  \bibinfo{volume}{9} (\bibinfo{year}{2012}), \bibinfo{pages}{439--448}.
\newblock


\bibitem[\protect\citeauthoryear{Lindvall, Rus, Shull, Zelkowitz, Donzelli,
  Memon, Basili, Costa, Tvedt, Hochstein, Asgari, Ackermann, and Pech}{Lindvall
  et~al\mbox{.}}{2005}]%
        {Lindvall05}
\bibfield{author}{\bibinfo{person}{M. Lindvall}, \bibinfo{person}{I. Rus},
  \bibinfo{person}{F. Shull}, \bibinfo{person}{M. Zelkowitz},
  \bibinfo{person}{P. Donzelli}, \bibinfo{person}{A. Memon},
  \bibinfo{person}{V. Basili}, \bibinfo{person}{P. Costa}, \bibinfo{person}{R.
  Tvedt}, \bibinfo{person}{L. Hochstein}, \bibinfo{person}{S. Asgari},
  \bibinfo{person}{C. Ackermann}, {and} \bibinfo{person}{D. Pech}.}
  \bibinfo{year}{2005}\natexlab{}.
\newblock \showarticletitle{An evolutionary testbed for software technology
  evaluation}.
\newblock \bibinfo{journal}{\emph{Innovations in Systems and Software
  Engineering}} \bibinfo{volume}{1}, \bibinfo{number}{1} (\bibinfo{date}{01
  Apr} \bibinfo{year}{2005}), \bibinfo{pages}{3--11}.
\newblock


\bibitem[\protect\citeauthoryear{Meng, Thain, Vyushkov, Wolf, and Woodard}{Meng
  et~al\mbox{.}}{2016}]%
        {meng2016conducting}
\bibfield{author}{\bibinfo{person}{H. Meng}, \bibinfo{person}{D. Thain},
  \bibinfo{person}{A. Vyushkov}, \bibinfo{person}{M. Wolf}, {and}
  \bibinfo{person}{A. Woodard}.} \bibinfo{year}{2016}\natexlab{}.
\newblock \showarticletitle{Conducting reproducible research with Umbrella:
  Tracking, creating, and preserving execution environments}. In
  \bibinfo{booktitle}{\emph{International Conference on e-Science}}.
  \bibinfo{pages}{91--100}.
\newblock


\bibitem[\protect\citeauthoryear{Miranda, Ferreira, de~Souza, Filho, and
  Singer}{Miranda et~al\mbox{.}}{2014}]%
        {DBLP:conf/icse/MirandaFSFS14}
\bibfield{author}{\bibinfo{person}{M. Miranda}, \bibinfo{person}{R. Ferreira},
  \bibinfo{person}{C.~R.~B. de Souza}, \bibinfo{person}{F.~M.~F. Filho}, {and}
  \bibinfo{person}{L. Singer}.} \bibinfo{year}{2014}\natexlab{}.
\newblock \showarticletitle{An exploratory study of the adoption of mobile
  development platforms by software engineers}. In
  \bibinfo{booktitle}{\emph{International Conference on Mobile Software
  Engineering and Systems}} \emph{(\bibinfo{series}{MOBILESoft '14})}.
  \bibinfo{pages}{50--53}.
\newblock


\bibitem[\protect\citeauthoryear{Morse}{Morse}{1991}]%
        {morse1991qualitative}
\bibfield{author}{\bibinfo{person}{J.~M. Morse}.}
  \bibinfo{year}{1991}\natexlab{}.
\newblock \showarticletitle{Qualitative nursing research: A free-for-all}.
\newblock \bibinfo{journal}{\emph{Qualitative nursing research: A contemporary
  dialogue}} (\bibinfo{year}{1991}), \bibinfo{pages}{14--22}.
\newblock


\bibitem[\protect\citeauthoryear{Murphy{-}Hill, Smith, Sadowski, Jaspan,
  Winter, Jorde, Knight, Trenk, and Gross}{Murphy{-}Hill et~al\mbox{.}}{2019}]%
        {DBLP:conf/icse/Murphy-HillSSJW19}
\bibfield{author}{\bibinfo{person}{E.~R. Murphy{-}Hill}, \bibinfo{person}{E.~K.
  Smith}, \bibinfo{person}{C. Sadowski}, \bibinfo{person}{C. Jaspan},
  \bibinfo{person}{C. Winter}, \bibinfo{person}{M. Jorde}, \bibinfo{person}{A.
  Knight}, \bibinfo{person}{A. Trenk}, {and} \bibinfo{person}{S. Gross}.}
  \bibinfo{year}{2019}\natexlab{}.
\newblock \showarticletitle{Do developers discover new tools on the toilet?}.
  In \bibinfo{booktitle}{\emph{International Conference on Software
  Engineering}} \emph{(\bibinfo{series}{ICSE '19})}. \bibinfo{pages}{465--475}.
\newblock


\bibitem[\protect\citeauthoryear{Narayanan and Shmatikov}{Narayanan and
  Shmatikov}{2008}]%
        {narayanan2008robust}
\bibfield{author}{\bibinfo{person}{A. Narayanan} {and} \bibinfo{person}{V.
  Shmatikov}.} \bibinfo{year}{2008}\natexlab{}.
\newblock \showarticletitle{Robust de-anonymization of large sparse datasets}.
  In \bibinfo{booktitle}{\emph{Symposium on Security and Privacy}}
  \emph{(\bibinfo{series}{SP '08})}. \bibinfo{pages}{111--125}.
\newblock


\bibitem[\protect\citeauthoryear{{NIH}}{{NIH}}{2003}]%
        {nih-sharing-policy}
\bibfield{author}{\bibinfo{person}{{NIH}}.} \bibinfo{year}{2003}\natexlab{}.
\newblock \bibinfo{title}{{NIH Data Sharing Policy and Implementation
  Guidance}}.
\newblock
\newblock
\urldef\tempurl%
\url{https://grants.nih.gov/grants/policy/data_sharing/data_sharing_guidance.htm}
\showURL{%
\tempurl}
\newblock
\shownote{[Date Accessed: July 21st 2020].}


\bibitem[\protect\citeauthoryear{{NSF}}{{NSF}}{2011}]%
        {nsf-sharing-policy}
\bibfield{author}{\bibinfo{person}{{NSF}}.} \bibinfo{year}{2011}\natexlab{}.
\newblock \bibinfo{title}{{Dissemination and Sharing of Research Results}}.
\newblock
\newblock
\urldef\tempurl%
\url{https://www.nsf.gov/bfa/dias/policy/dmp.jsp}
\showURL{%
\tempurl}
\newblock
\shownote{[Date Accessed: July 21st 2020].}


\bibitem[\protect\citeauthoryear{O'Neill, Pouder, and Buchholtz}{O'Neill
  et~al\mbox{.}}{1998}]%
        {o1998patterns}
\bibfield{author}{\bibinfo{person}{H.~M. O'Neill}, \bibinfo{person}{R.~W.
  Pouder}, {and} \bibinfo{person}{A.~K. Buchholtz}.}
  \bibinfo{year}{1998}\natexlab{}.
\newblock \showarticletitle{Patterns in the diffusion of strategies across
  organizations: Insights from the innovation diffusion literature}.
\newblock \bibinfo{journal}{\emph{Academy of Management Review}}
  \bibinfo{volume}{23}, \bibinfo{number}{1} (\bibinfo{year}{1998}),
  \bibinfo{pages}{98--114}.
\newblock


\bibitem[\protect\citeauthoryear{{Oracle}}{{Oracle}}{2021}]%
        {virtualbox}
\bibfield{author}{\bibinfo{person}{{Oracle}}.} \bibinfo{year}{2021}\natexlab{}.
\newblock \bibinfo{title}{{VirtualBox}}.
\newblock
\newblock
\urldef\tempurl%
\url{https://www.virtualbox.org}
\showURL{%
\tempurl}
\newblock
\shownote{[Date Accessed: March \nth{13} 2021].}


\bibitem[\protect\citeauthoryear{Patterson, Snyder, and Ullman}{Patterson
  et~al\mbox{.}}{1999}]%
        {cra-best-practices}
\bibfield{author}{\bibinfo{person}{D. Patterson}, \bibinfo{person}{L. Snyder},
  {and} \bibinfo{person}{J. Ullman}.} \bibinfo{year}{1999}\natexlab{}.
\newblock \showarticletitle{{Evaluating Computer Scientists and Engineers For
  Promotion and Tenure}}.
\newblock \bibinfo{journal}{\emph{Computing Research News}}
  (\bibinfo{date}{August} \bibinfo{year}{1999}).
\newblock


\bibitem[\protect\citeauthoryear{Premkumar, Ramamurthy, and
  Nilakanta}{Premkumar et~al\mbox{.}}{1994}]%
        {premkumar1994implementation}
\bibfield{author}{\bibinfo{person}{G. Premkumar}, \bibinfo{person}{K.
  Ramamurthy}, {and} \bibinfo{person}{S. Nilakanta}.}
  \bibinfo{year}{1994}\natexlab{}.
\newblock \showarticletitle{Implementation of electronic data interchange: an
  innovation diffusion perspective}.
\newblock \bibinfo{journal}{\emph{Journal of Management Information Systems}}
  \bibinfo{volume}{11}, \bibinfo{number}{2} (\bibinfo{year}{1994}),
  \bibinfo{pages}{157--186}.
\newblock


\bibitem[\protect\citeauthoryear{{PyPA}}{{PyPA}}{2021}]%
        {pip}
\bibfield{author}{\bibinfo{person}{{PyPA}}.} \bibinfo{year}{2021}\natexlab{}.
\newblock \bibinfo{title}{{pip -- The Python Package Installer}}.
\newblock
\newblock
\urldef\tempurl%
\url{https://pip.pypa.io/en/stable}
\showURL{%
\tempurl}
\newblock
\shownote{[Date Accessed: March \nth{13} 2021].}


\bibitem[\protect\citeauthoryear{QEMU}{QEMU}{2021}]%
        {qemu}
\bibfield{author}{\bibinfo{person}{QEMU}.} \bibinfo{year}{2021}\natexlab{}.
\newblock \bibinfo{title}{QEMU}.
\newblock
\newblock
\urldef\tempurl%
\url{https://www.qemu.org}
\showURL{%
\tempurl}
\newblock
\shownote{[Date Accessed: March \nth{13} 2021].}


\bibitem[\protect\citeauthoryear{{RedHat}}{{RedHat}}{2021}]%
        {podman}
\bibfield{author}{\bibinfo{person}{{RedHat}}.} \bibinfo{year}{2021}\natexlab{}.
\newblock \bibinfo{title}{Podman}.
\newblock
\newblock
\urldef\tempurl%
\url{https://github.com/containers/podman}
\showURL{%
\tempurl}
\newblock
\shownote{[Date Accessed: March \nth{13} 2021].}


\bibitem[\protect\citeauthoryear{Rogers}{Rogers}{2010}]%
        {rogers2010diffusion}
\bibfield{author}{\bibinfo{person}{E.~M. Rogers}.}
  \bibinfo{year}{2010}\natexlab{}.
\newblock \bibinfo{booktitle}{\emph{Diffusion of innovations}}.
\newblock \bibinfo{publisher}{Simon and Schuster}.
\newblock


\bibitem[\protect\citeauthoryear{{Rougier}, {Hinsen}, {Alexandre}, {Arildsen},
  {Barba}, {Benureau}, {Brown}, de~{Buyl}, {Caglayan}, {Davison}, {Delsuc},
  {Detorakis}, {Diem}, {Drix}, {Enel}, {Girard}, {Guest}, {Hall}, {Henriques},
  {Hinaut}, {Jaron}, {Khamassi}, {Klein}, {Manninen}, {Marchesi}, {McGlinn},
  {Metzner}, {Petchey}, {Plesser}, {Poisot}, {Ram}, {Ram}, {Roesch}, {Rossant},
  {Rostami}, {Shifman}, {Stachelek}, {Stimberg}, {Stollmeier}, {Vaggi},
  {Viejo}, {Vitay}, {Vostinar}, {Yurchak}, and {Zito}}{{Rougier}
  et~al\mbox{.}}{2017}]%
        {Rougier2017}
\bibfield{author}{\bibinfo{person}{N.~P. {Rougier}}, \bibinfo{person}{K.
  {Hinsen}}, \bibinfo{person}{F. {Alexandre}}, \bibinfo{person}{T. {Arildsen}},
  \bibinfo{person}{L.~A. {Barba}}, \bibinfo{person}{F.~C.~Y. {Benureau}},
  \bibinfo{person}{C.~Titus {Brown}}, \bibinfo{person}{P. de {Buyl}},
  \bibinfo{person}{O. {Caglayan}}, \bibinfo{person}{A.~P. {Davison}},
  \bibinfo{person}{M. {Delsuc}}, \bibinfo{person}{G. {Detorakis}},
  \bibinfo{person}{A.~K. {Diem}}, \bibinfo{person}{D. {Drix}},
  \bibinfo{person}{P. {Enel}}, \bibinfo{person}{B. {Girard}},
  \bibinfo{person}{O. {Guest}}, \bibinfo{person}{M.~G. {Hall}},
  \bibinfo{person}{R.~N. {Henriques}}, \bibinfo{person}{X. {Hinaut}},
  \bibinfo{person}{K.~S. {Jaron}}, \bibinfo{person}{M. {Khamassi}},
  \bibinfo{person}{A. {Klein}}, \bibinfo{person}{T. {Manninen}},
  \bibinfo{person}{P. {Marchesi}}, \bibinfo{person}{D. {McGlinn}},
  \bibinfo{person}{C. {Metzner}}, \bibinfo{person}{O.~L. {Petchey}},
  \bibinfo{person}{H.~E. {Plesser}}, \bibinfo{person}{T. {Poisot}},
  \bibinfo{person}{K. {Ram}}, \bibinfo{person}{Y. {Ram}},
  \bibinfo{person}{E.~B. {Roesch}}, \bibinfo{person}{C. {Rossant}},
  \bibinfo{person}{V. {Rostami}}, \bibinfo{person}{A. {Shifman}},
  \bibinfo{person}{J. {Stachelek}}, \bibinfo{person}{M. {Stimberg}},
  \bibinfo{person}{F. {Stollmeier}}, \bibinfo{person}{F. {Vaggi}},
  \bibinfo{person}{G. {Viejo}}, \bibinfo{person}{J. {Vitay}},
  \bibinfo{person}{A.~E. {Vostinar}}, \bibinfo{person}{R. {Yurchak}}, {and}
  \bibinfo{person}{T. {Zito}}.} \bibinfo{year}{2017}\natexlab{}.
\newblock \showarticletitle{{Sustainable computational science: the ReScience
  initiative}}.
\newblock \bibinfo{journal}{\emph{PeerJ Computer Science}}  \bibinfo{volume}{3}
  (\bibinfo{year}{2017}), \bibinfo{pages}{1--8}.
\newblock
Issue 12.


\bibitem[\protect\citeauthoryear{Salda{\~n}a}{Salda{\~n}a}{2015}]%
        {saldana2015coding}
\bibfield{author}{\bibinfo{person}{J. Salda{\~n}a}.}
  \bibinfo{year}{2015}\natexlab{}.
\newblock \bibinfo{booktitle}{\emph{The coding manual for qualitative
  researchers}}.
\newblock \bibinfo{publisher}{Sage}.
\newblock


\bibitem[\protect\citeauthoryear{Shull, Basili, Carver, Maldonado, Travassos,
  Mendonca, and Fabbri}{Shull et~al\mbox{.}}{2002}]%
        {Shull02}
\bibfield{author}{\bibinfo{person}{F. Shull}, \bibinfo{person}{V. Basili},
  \bibinfo{person}{J. Carver}, \bibinfo{person}{J.~C. Maldonado},
  \bibinfo{person}{G.~H. Travassos}, \bibinfo{person}{M. Mendonca}, {and}
  \bibinfo{person}{S. Fabbri}.} \bibinfo{year}{2002}\natexlab{}.
\newblock \showarticletitle{Replicating software engineering experiments:
  addressing the tacit knowledge problem}. In
  \bibinfo{booktitle}{\emph{International Symposium on Empirical Software
  Engineering}} \emph{(\bibinfo{series}{ESEM '02})}. \bibinfo{pages}{7--16}.
\newblock


\bibitem[\protect\citeauthoryear{Shull, Carver, Vegas, and Juristo}{Shull
  et~al\mbox{.}}{2008}]%
        {Shull08}
\bibfield{author}{\bibinfo{person}{F.~J. Shull}, \bibinfo{person}{J.~C.
  Carver}, \bibinfo{person}{S. Vegas}, {and} \bibinfo{person}{N. Juristo}.}
  \bibinfo{year}{2008}\natexlab{}.
\newblock \showarticletitle{{The role of replications in Empirical Software
  Engineering}}.
\newblock \bibinfo{journal}{\emph{Empirical Software Engineering}}
  \bibinfo{volume}{13}, \bibinfo{number}{2} (\bibinfo{date}{01 Apr}
  \bibinfo{year}{2008}), \bibinfo{pages}{211--218}.
\newblock


\bibitem[\protect\citeauthoryear{Silverman and Marvasti}{Silverman and
  Marvasti}{2008}]%
        {silverman2008doing}
\bibfield{author}{\bibinfo{person}{D. Silverman} {and} \bibinfo{person}{A.
  Marvasti}.} \bibinfo{year}{2008}\natexlab{}.
\newblock \bibinfo{booktitle}{\emph{Doing qualitative research: A comprehensive
  guide}}.
\newblock \bibinfo{publisher}{Sage}.
\newblock


\bibitem[\protect\citeauthoryear{Smith, Katz, and Niemeyer}{Smith
  et~al\mbox{.}}{2016}]%
        {Smith16}
\bibfield{author}{\bibinfo{person}{A.~M. Smith}, \bibinfo{person}{D.~S. Katz},
  {and} \bibinfo{person}{K.~E. Niemeyer}.} \bibinfo{year}{2016}\natexlab{}.
\newblock \showarticletitle{{Software Citation Principles}}.
\newblock \bibinfo{journal}{\emph{PeerJ Computer Science}}  \bibinfo{volume}{2}
  (\bibinfo{year}{2016}), \bibinfo{pages}{e86}.
\newblock


\bibitem[\protect\citeauthoryear{Stodden}{Stodden}{2009a}]%
        {StoddenIJCLP09}
\bibfield{author}{\bibinfo{person}{V. Stodden}.}
  \bibinfo{year}{2009}\natexlab{a}.
\newblock \showarticletitle{Enabling Reproducible Research: Open Licensing for
  Scientific Innovation}.
\newblock \bibinfo{journal}{\emph{International Journal of Communications Law
  and Policy}}  \bibinfo{volume}{13} (\bibinfo{year}{2009}).
\newblock


\bibitem[\protect\citeauthoryear{Stodden}{Stodden}{2009b}]%
        {StoddenCSE09}
\bibfield{author}{\bibinfo{person}{V. Stodden}.}
  \bibinfo{year}{2009}\natexlab{b}.
\newblock \showarticletitle{The Legal Framework for Reproducible Scientific
  Research: Licensing and Copyright}.
\newblock \bibinfo{journal}{\emph{Computing in Science \& Engineering}}
  \bibinfo{volume}{11}, \bibinfo{number}{1} (\bibinfo{year}{2009}),
  \bibinfo{pages}{35--40}.
\newblock


\bibitem[\protect\citeauthoryear{Stodden, Leisch, and Peng}{Stodden
  et~al\mbox{.}}{2014}]%
        {stodden2014implementing}
\bibfield{author}{\bibinfo{person}{V. Stodden}, \bibinfo{person}{F. Leisch},
  {and} \bibinfo{person}{R.~D. Peng}.} \bibinfo{year}{2014}\natexlab{}.
\newblock \bibinfo{booktitle}{\emph{Implementing reproducible research}}.
\newblock \bibinfo{publisher}{CRC Press}.
\newblock


\bibitem[\protect\citeauthoryear{Teshima, Cleary, and Singer}{Teshima
  et~al\mbox{.}}{2013}]%
        {DBLP:conf/msr/GomezCS13}
\bibfield{author}{\bibinfo{person}{C.~G. Teshima}, \bibinfo{person}{B. Cleary},
  {and} \bibinfo{person}{L. Singer}.} \bibinfo{year}{2013}\natexlab{}.
\newblock \showarticletitle{A study of innovation diffusion through link
  sharing on stack overflow}. In \bibinfo{booktitle}{\emph{Working Conference
  on Mining Software Repositories}} \emph{(\bibinfo{series}{MSR '13})}.
  \bibinfo{pages}{81--84}.
\newblock


\bibitem[\protect\citeauthoryear{Timperley, Stepney, and {Le Goues}}{Timperley
  et~al\mbox{.}}{2018}]%
        {bugzoo}
\bibfield{author}{\bibinfo{person}{C.~S. Timperley}, \bibinfo{person}{S.
  Stepney}, {and} \bibinfo{person}{C. {Le Goues}}.}
  \bibinfo{year}{2018}\natexlab{}.
\newblock \showarticletitle{BugZoo: A platform for studying software bugs}. In
  \bibinfo{booktitle}{\emph{International Conference on Software Engineering:
  Companion Proceeedings}}. \bibinfo{pages}{446--447}.
\newblock


\bibitem[\protect\citeauthoryear{Wolfe}{Wolfe}{1994}]%
        {wolfe1994organizational}
\bibfield{author}{\bibinfo{person}{R.~A. Wolfe}.}
  \bibinfo{year}{1994}\natexlab{}.
\newblock \showarticletitle{Organizational innovation: Review, critique and
  suggested research directions}.
\newblock \bibinfo{journal}{\emph{Journal of Management Studies}}
  \bibinfo{volume}{31}, \bibinfo{number}{3} (\bibinfo{year}{1994}),
  \bibinfo{pages}{405--431}.
\newblock


\bibitem[\protect\citeauthoryear{Wright, Palmar, and Kavanaugh}{Wright
  et~al\mbox{.}}{1995}]%
        {wright1995importance}
\bibfield{author}{\bibinfo{person}{R.~E. Wright}, \bibinfo{person}{J.~C.
  Palmar}, {and} \bibinfo{person}{D.~C. Kavanaugh}.}
  \bibinfo{year}{1995}\natexlab{}.
\newblock \showarticletitle{The importance of promoting stakeholder acceptance
  of educational innovations}.
\newblock \bibinfo{journal}{\emph{Education}} \bibinfo{volume}{115},
  \bibinfo{number}{4} (\bibinfo{year}{1995}), \bibinfo{pages}{628--633}.
\newblock


\bibitem[\protect\citeauthoryear{Zerhouni}{Zerhouni}{2003}]%
        {zerhouni2003nih}
\bibfield{author}{\bibinfo{person}{E. Zerhouni}.}
  \bibinfo{year}{2003}\natexlab{}.
\newblock \bibinfo{title}{The NIH roadmap}.
\newblock
\newblock


\end{thebibliography}
\end{document}